\documentclass{aa}  

\usepackage{graphicx}
\usepackage{float}
\usepackage{tabularx}
\usepackage{wasysym}
\usepackage{soul}
\usepackage{natbib}
\usepackage[colorlinks=true,linkcolor=blue,citecolor=blue]{hyperref}%
\usepackage[dvipsnames, svgnames, table, xcdraw]{xcolor}
\definecolor{skobeloff}{rgb}{0.0, 0.48, 0.45}
\DeclareUnicodeCharacter{0301}{\'{e}}
 
\bibpunct{(}{)}{;}{a}{}{,} 
\usepackage{txfonts}
\begin{document}

   \title{AGN feeding along a one-armed spiral in NGC\,4593}

   \subtitle{A study using ALMA CO(2-1) observations}

        \author{K. Kianfar 
          \inst{1,}\inst{2}\thanks{\email{khashayar@ita.br}}
          \and
          P. Andreani\inst{1}
          \and J.A. Fernández-Ontiveros\inst{3}
          \and F. Combes \inst{4}
          \and L. Spinoglio \inst{5}
          \and E. Hatziminaoglou\inst{1,}\inst{6,}\inst{7}
          \and C. Ricci\inst{8, }\inst{9}
          \and A. Bewketu-Belete \inst{10}
          \and M. Imanishi \inst{11, }\inst{12}
          \and M. Pereira-Santaella \inst{13}
          \and R. Slater \inst{14}
          \and M. Malheiro\inst{2}\thanks{Deceased: It is with profound sadness we note the passing of our esteemed colleague, M. Malheiro, who made significant contributions to this work.}}

\institute{ European Southern Observatory (ESO), Karl-Schwarzschild-Strasse 2, 85748 Garching bei München, Germany
        \and
        Physics Department, Aeronautics Institute
    of Technology - ITA, Praça Marechal Eduardo
      Gomes, 50, São José dos Campos, 12228-900,
    São Paulo, Brazil
         \and        
        Centro de Estudios de Física del Cosmos de Aragón (CEFCA), Plaza San Juan 1, 44001 Teruel, Spain
        \and
        LERMA, Observatoire de Paris, Collége de France, PSL University, CNRS, Sorbonne University, Paris, France
        \and
        INAF-Istituto di Astrofisica e Planetologia Spaziali, via del Fosso del
        Cavaliere, 100, 00100 Roma, Italy
        \and
        Instituto de Astrofísica de Canarias (IAC), E-38205 La Laguna, Tenerife, Spain
        \and
        Universidad de La Laguna, Dpto. Astrofísica, E-38206 La Laguna, Tenerife, Spain
        \and
        Instituto de Estudios Astrofísicos, Facultad de Ingenier\'ia y Ciencias, Universidad Diego Portales, Av. Ej\'ercito Libertador 441, Santiago, Chile
        \and
        Kavli Institute for Astronomy and Astrophysics, Peking University, Beijing 100871, China   
        \and
        Department of Space, Earth \& Environment, Chalmers University of Technology, SE-412 96 Gothenburg, Sweden
        \and 
        National Astronomical Observatory of Japan, National Institutes of Natural Sciences (NINS), 2-21-1 Osawa, Mitaka, Tokyo 181-8588, Japan 
        \and
        Department of Astronomy, School of Science, Graduate University for Advanced Studies (SOKENDAI), Mitaka, Tokyo 181-8588, Japan 
        \and 
        Instituto de Física Fundamental, CSIC, Calle Serrano 123, E-28006 Madrid, Spain
        \and
        Departamento de Tecnologías Industriales, Facultad de Ingeniería, Universidad de Talca, Los Niches km 1, Curicó, Chile
             }
   \date{Received...; accepted...}

 
\abstract
{We investigate active galactic nuclei (AGN) feeding through the molecular gas (CO(2-1) emission) properties of the local Seyfert 1 galaxy NGC\,4593, using Atacama Large Millimeter Array (ALMA) observations and other multi-wavelength data. }
{Our study aims to understand the interplay between the AGN and the interstellar medium (ISM) in this galaxy, examining the role of the AGN in steering gas dynamics within its host galaxy, evaluating the energy injected into the ISM, and determining whether gas is inflowing or outflowing from the galaxy.}
{After reducing the ALMA CO(2-1) images, we employed two models, \textsc{3D-Barolo} and \textsc{discFit}, to construct a disc model and fit its emission to the ALMA data. Additionally, we used photometric data to build a spectral energy distribution (SED) and apply the CIGALE code to derive key physical properties of the AGN and its host.
}
{Our analysis reveals a complex interplay within NGC\,4593, including a clear rotational pattern, the influence of a non-axisymmetric bar potential, and a central molecular zone (CMZ)-like ring. We observe an outflow of CO(2-1) gas along the minor axis, at a distance of $\sim$ 220 pc from the nucleus. The total molecular gas mass is estimated to be $1 - 5 \times 10^8 \, M_{\odot}$, with non-circular motions contributing $10\%$. Our SED analysis indicates an AGN fraction of 0.88 and a star formation rate (SFR) of 0.42 $M_{\odot}\,\text{yr}^{-1}$.
}
{These findings highlight the complex dynamics in the centre of NGC\,4593, which are significantly influenced by the presence of the AGN. The overall physical properties of this system suggest that the AGN has a substantial impact on the evolution of NGC\,4593.}


\maketitle
%

\section{Introduction}

The role of active galactic nuclei (AGN) in shaping the evolution of galaxies across the Universe is not well understood. AGNs are powered by the accretion of matter into a supermassive black hole (SMBH) in the galaxy's core, and in many cases, their luminosity outshines that of the host galaxy \citep{kauffmann2000unified, heckman2004present, hopkins2005black}.

Observational and theoretical studies in the last decades have highlighted a possible link between the redshift evolution of black hole (BH) accretion and the star formation rate in galaxies, which may result from feeding and feedback processes \citep{silk1998quasars, fabian2012observational}. The relative influence of matter accretion (feeding) versus feedback processes, such as energy released in the form of radiation, outflows, and jets, likely interacts with the surrounding medium, preventing the gas from cooling and condensing into stars. AGN feedback can also have positive effects, such as jet-induced star formation \citep[e.g.,][]{combes2017agn}. These processes are observed in various gas phases, including molecular, atomic, and ionised, and their interaction with host galaxies remains a topic of ongoing research \citep[e.g.,][]{cicone2014massive, hatziminaoglou2010hermes, morganti2017many, fiore2017agn, tombesi2015wind}.

Theoretical works suggest that even a tiny fraction of the vast energy produced by AGN can have a significant impact on the growth and evolution of galaxies, highlighting the importance of further research in this area \citep[e.g.,][]{mcnamara2007heating, gitti2012evidence, gaspari2020linking, schawinski2007observational}. Massive molecular outflows are observed in many galaxies and are generally interpreted as the long-sought feedback mechanism. However, the exact properties and timescales of this feedback are still not fully understood \citep{morganti2017many}. Several studies suggest that the observed outflows do not suppress star formation in their host galaxies \citep[e.g.,][]{feruglio2010quasar, harrison2013impact}. The relationship between AGN feedback and star formation is indeed complex, suggesting a potential interplay between these two processes \citep[e.g.,][]{harrison2017impact, girdhar2022quasar}. 

AGN radiation is often 'hidden' by large amounts of dust and gas, which prevents the study of the activities in the central part of galaxies where the interaction between AGN and host galaxy takes place \citep{hickox2018obscured}. Dust and gas absorb and scatter radiation from the central AGN, thereby attenuating the radiation intensity, and the dust grains mainly absorb the FUV radiation and re-emit it mostly at far-infrared (FIR) wavelengths. This reemission in the IR drastically affects the observable parameters of the AGN \citep{pier1992infrared}, and polarising its light provides important knowledge on the AGN structure and shape \citep{antonucci1985spectropolarimetry}. Studies demonstrate that radiative feedback from AGNs confirms that radiation pressure on dusty gas is the primary mechanism regulating the obscuration and, consequently, the visibility of AGN activities \citep{hensley2014grain, venanzi2020role, arakawa2022radiation, ishibashi2015agn, ricci2017close, ricci2022bass}. The AGN feedback effectively clears out the obscuring dust and gas within the BH, impacting its interaction with the host galaxy. The study by \citet{hopkins2012stellar} suggests that radiation pressure on dust grains is a significant, yet not exclusive, driver of galactic superwinds. In a specific case, \citet{contursi2013spectroscopic} observed in the outflow of M\,82 that dust is kinematically decoupled from gas, indicating that dust grains move slower than ionised and molecular gas and may fall back into the galaxy disc.

Carbon monoxide (CO) is a pivotal tracer for cold molecular gas in galaxies, detectable with submillimetre and radio telescopes \citep[e.g.,][]{bolatto2013co}. The several rotational transitions of the CO molecules provide insights into the molecular gas temperature and density \citep[e.g.,][]{weiss2005multiple, papadopoulos2010co, kamenetzky2018recovering, andreani2018extreme}. Several ALMA observations of galaxies have particularly focused on the dynamics of the gas around AGN \citep{veilleux2020cool, esposito2022agn}, highlighting the influence of AGN emission on the molecular gas properties. The exact nature of this influence, especially on star-forming gas, can be localised and complex \citep{rosario2019accreting}. Multi-transition CO data modelling further refines our understanding of the AGN impact on molecular gas \citep{van2010black, fluetsch2019cold}.

High-resolution CO observations are crucial to constrain the spatial and velocity dynamics of molecular gas \citep{sun2018cloud}. Through moment maps, we derive the properties of the gas dynamics, with zeroth, first, and second moment maps detailing the CO emission distribution, line-of-sight velocity, and velocity dispersion, respectively \citep[e.g.,][]{davis2013atlas3d}. Different studies examine the relationship between CO molecular gas and obscured AGN, highlighting the intricate interplay with dust. For instance, ALMA observations of dust-obscured quasars at high redshifts unveil molecular gas outflows and feedback mechanisms, suggesting a potential evolutionary sequence between star-forming galaxies, ultra-luminous infrared galaxies, and quasars \citep{spingola2019erratum, brusa2017molecular}. However, this correlation may primarily be driven by underlying factors such as stellar mass and SFR. Interestingly, when controlling for star formation, \citet{koss2021bat} find no significant differences in molecular gas properties between AGN-hosting galaxies and inactive galaxies, indicating that AGN feedback might not have a discernible impact on the molecular gas reservoirs.

Molecular outflows are usually identified in high spectral resolution data cubes as high-velocity gas deviating from a regular gas rotation \citep[e.g.,][]{cicone2014massive}. In a comprehensive study, \citet{lutz2020molecular} discuss methods for detecting such outflows, emphasising the intermittent activity of AGN as a potential driver \citep[e.g.,][]{newman2012sins}. Additionally, \citet{stuber2021frequency} find that central molecular outflows are relatively common in nearby spiral galaxies, especially those with higher stellar mass and SFR. These outflows often link to AGN and bars, but they seem to have a limited effect on suppressing star formation due to their low mass-loading factors.

In this paper, we report a study of the galaxy NGC\,4593, which belongs to the Twelve-micron WInd STatistics (TWIST) sample, a CO(2–1) molecular gas survey of 41 galaxies drawn from the 12-micron sample \citep{rush1993extended}. Other studies \citep{Juan+20,Asnakew} have reported results derived from four galaxies in the same sample.

\begin{figure*}[!htbp]
\centering
    \includegraphics[width=5.66cm]{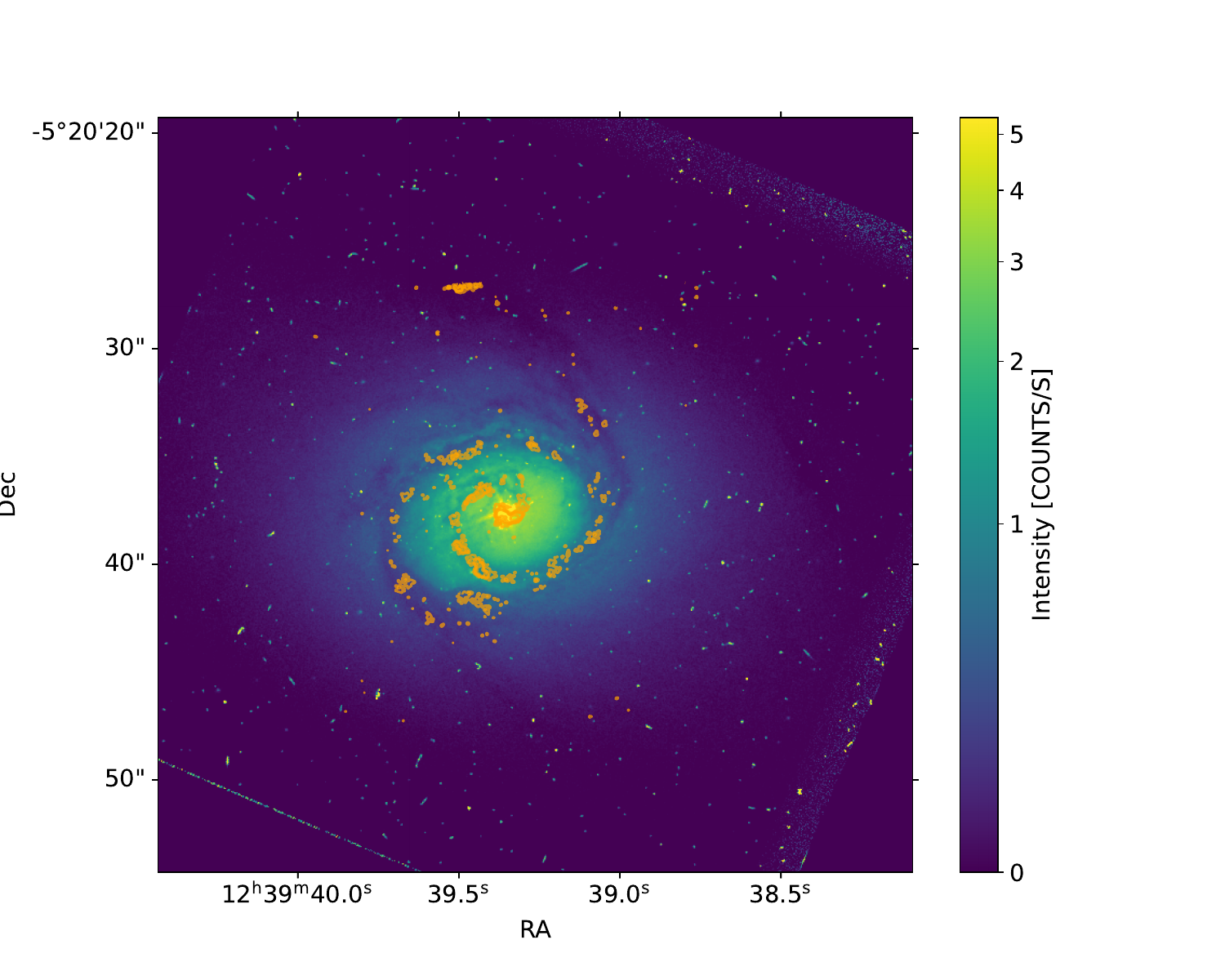}
    \includegraphics[width=5.66cm]{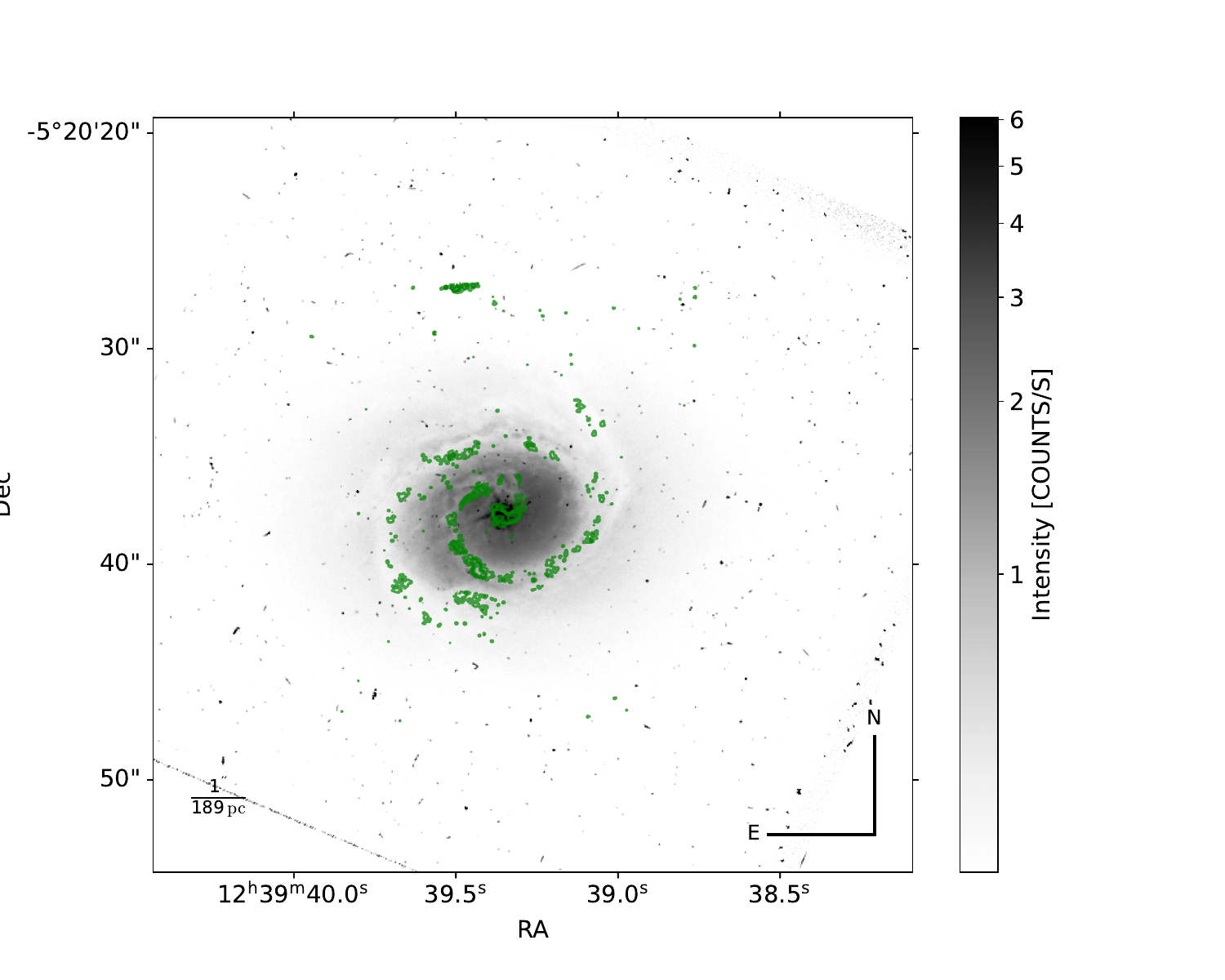}
    \includegraphics[width=5.66cm]{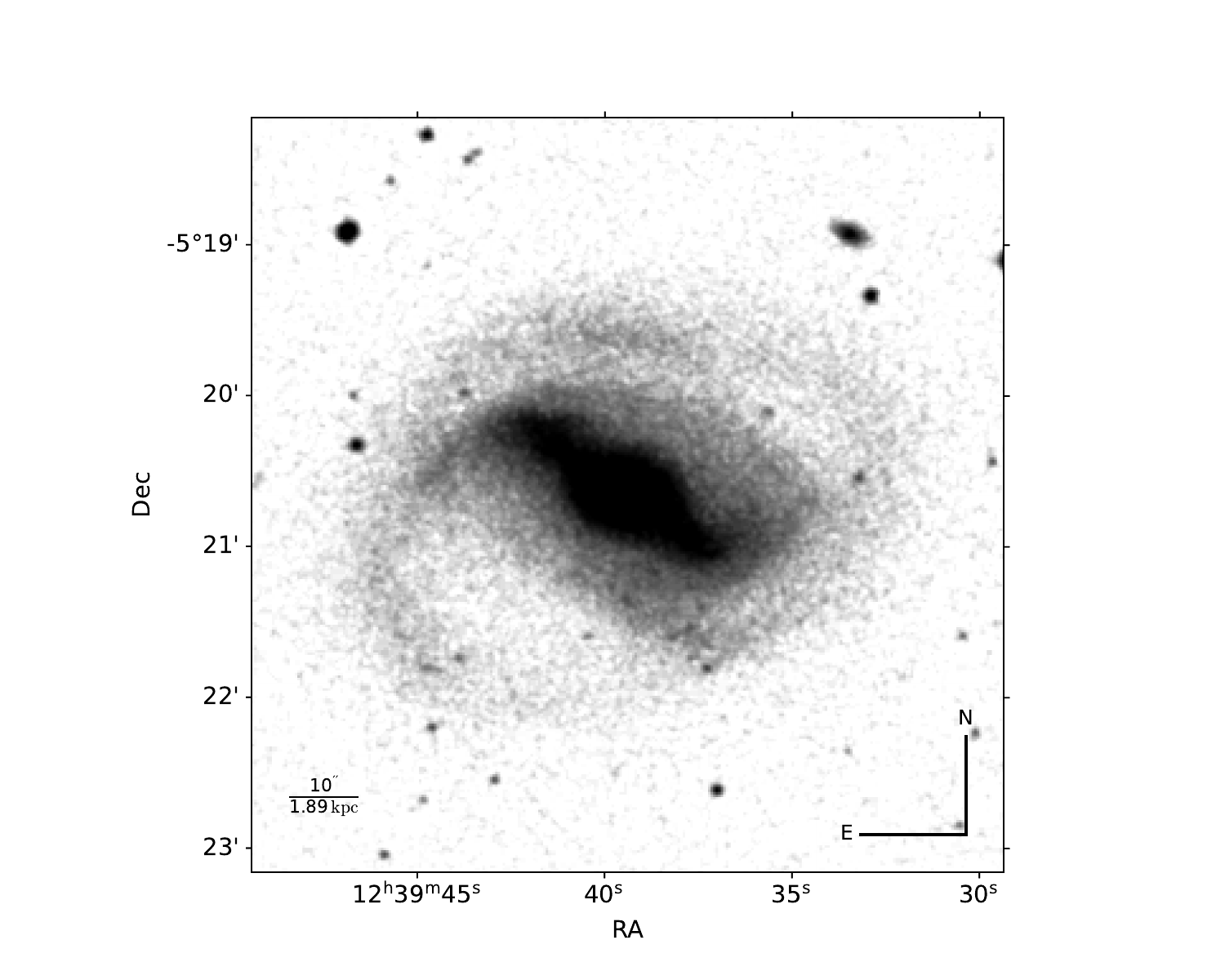}
    \caption{Multi-wavelength observations of NGC\,4593. Left: Background merged image from \textit{Hubble} Space Telescope (HST) observations, combining data from the Wide Field Planetary Camera 2 (WFPC2) using the F606W filter and the F547\,M filter. The squared zoomed region highlights the inner galaxy, showing the AGN at its centre. Middle: Zoomed region of the HST image with contours from the ALMA CO (J=2-1) emission line moment-0 map, illustrating the distribution and intensity of CO emission with contour levels of $Contours \, = \, [0.25, 0.35, 0.4, 0.5, 1, 2, 4, 6]$. Right: Digitized Sky Survey (DSS) image of the galaxy (Image credit: DSS, STScI).}
    \label{fig:hst}
\end{figure*}

NGC\,4593 is a face-on spiral galaxy in the southern sky ($\alpha = 12^h\,39^m\,39.4^s, \, \delta = -5^\circ\, 20'\, 39''$) located at $D \approx 39 \pm 4$ Mpc \citep{marinova2007characterizing}. This galaxy is a well-defined barred spiral with nuclear, inner, and outer rings, classified morphologically as (R)SB(rs)b \citep{deVaucouleurs1991}. It contains a prominent bar that likely drives gas inflow toward the nucleus. NGC\,4593 has a nuclear dust ring linked to radial dust lanes in the galaxy's bar, indicative of gas inflow \citep[e.g.,][]{wu2021morphological, hunt2008molecular, laine2001ngc}. The accretion disc size in NGC\,4593 exceeds that of the broad-line region, hinting at a complex central engine structure \citep[e.g.,][]{cackett2019comparison}. NGC\,4593 is a well-studied Seyfert 1 galaxy, exhibiting polarised continuum emission, X-ray, UV, and radio variability, as well as time lags between FUV and NUV bands, and between $0.3-0.5 \, \rm keV$ vs. $0.8 - 1.5 \, \rm keV$ and $0.3-0.5 \, \rm keV$ vs. $4.0-10.0\, \rm keV$ energy bands \citep{sriram2009energy, kumari2023contrasting, kammoun2021modelling}.

The density profile of NGC\,4593's circumnuclear medium aligns with a power-law distribution, indicating a continuous gas distribution around the central BH \citep{wang2022density}. In Table~\ref{tab:table1}, we summarise the observed properties and measured quantities for NGC\,4593, providing an overview of the galaxy's physical characteristics, which are used in this work.

The main goal of this paper is to study the molecular gas properties in the central part of NGC\,4593 using ALMA observations of the CO(2-1) transition. Using these observations (Section~\ref{obs}), we model the data to determine the kinematics and morphology of the gas (Section~\ref{res}). To acquire a better understanding of the interaction between the molecular gas and the AGN, we further enhance our study by using observations at other wavelengths to create a spectral energy distribution (SED). This approach provides a more holistic perspective of the galaxy, allowing for a better understanding of the influence of AGN activity on its host galaxy. We discuss how our findings contribute to the ongoing debate over the mechanisms driving AGN feeding, feedback, and dust emission, as well as their impact on galaxy evolution, in Section~\ref{Discussion}.
\begin{table}[h!]
\centering
\caption{Observed properties of NGC\,4593}
\label{tab:table1}
\begin{tabular}{l l l}
\hline\hline
Parameter & Value & Ref. \\
\hline
Name & NGC\,4593 & (1) \\
RA & 12:39:39.43 & (1) \\
Dec & $-5$:20:39.3 & (1) \\
\textit{Hubble} class & (R)SB(rs)b & (2) \\
$z$ & 0.0083 & (1) \\
$D_L$ & 39 Mpc & (1) \\
Type & S1 & (3) \\
$F_{12}$ & 0.47 Jy & (4) \\
$\log(M_*/M_{\odot})$ & 10.6--11.01 & (5) \\
SFR & $0.19$--$1.12$ $M_{\odot}$\,yr$^{-1}$ & (6) \\
$\log(L_{\mathrm{IR}})$ & 43.431 & (7) \\
PA & $-59.3^{\circ}$ & (8) \\
$V_{\mathrm{sys}}$ & 2492 km\,s$^{-1}$ & (2) \\  
$i$ & $47.4^{\circ}$ & (8) \\
$F_{\mathrm{PAH2}}$ & $287.2 \pm 7.4$ mJy & (9) \\
$F_{\mathrm{mid\text{-}IR}}$ & $352 \pm 8$ mJy & (10) \\
\hline
\end{tabular}
\tablefoot{
List of observed properties includes coordinates, redshift, luminosity distance, star formation rate (SFR), infrared luminosity, position angle (PA), systemic velocity, and inclination.
}
\tablebib{
(1)~NASA/IPAC Extragalactic Database (NED); (2)~\citet{de2013third}; (3)~\citet{contini1998starbursts}; (4)~IRAS $12\,\mu\mathrm{m}$ flux;
(5)~\citet{schombert2019mass}; (6)~\citet{fernandez2016far,mordini2021calibration}; (7)~\citet{spinoglio1995multiwavelength}; (8)~Value derived in this work with 3D-Barolo model;
(9)~\citet{horst2006small,horst2008mid,horst2009mid}; (10)~\citet{asmus2014subarcsecond}.
}
\end{table}

\section{Observations}\label{obs}
\begin{table*}[h!]
\centering
\caption{ALMA observations of NGC\,4593}
\label{tab:table2}
\tiny{
\begin{tabular}{l l c c c c c c c c c}

\hline\hline
Band & Emission & Frequency & BMAJ & BMIN & BPA & RMS & $\rm V_{sys}$ & Velocity width & Velocity range & Velocity res. \\
     &           & (GHz)     & (arcsec) & (arcsec) & ($^{\circ}$) & (Jy beam$^{-1}$ km\,s$^{-1}$) & (km\,s$^{-1}$) & (km\,s$^{-1}$) & (km\,s$^{-1}$) & (km\,s$^{-1}$) \\
\hline
6 (mosaic) & CO(2--1) & 228.628 & 0.232 & 0.183 & $-59.3$ & $7 \times 10^{-4}$ & 2491 & 310 & 1990 & 2.4 \\
\hline
\end{tabular}
}
\tablefoot{
Details of Band 6 (mosaic) observations of NGC\,4593 (Project ID: 2017.1.00236.S). Listed parameters include the ALMA Band, target emission line, central frequency, and synthesised beam properties (BMAJ, BMIN, and BPA). The RMS column indicates the noise level, $\rm V_{sys}$ is the systemic velocity, and velocity width represents the observed line width. Velocity range denotes the data cube's velocity coverage relative to $\rm V_{sys}$, and velocity resolution specifies the observation's resolution.
}
\end{table*}

\subsection{ALMA mm/sub-mm interferometry}

NGC\,4593 was observed by the Atacama Large Millimeter/submillimeter Array (ALMA) on January 4th, 2018 (Cycle 5, PI Matthew A. Malkan, Program ID: 2017.1.00236.S). The spectral setup was optimised for the CO(2-1) transition line in band 6, at a rest frequency of 230.5380 GHz. The synthesised beam size was $0.029^{\prime\prime} \times 0.027^{\prime\prime}$, and the rms sensitivity was $0.36$ mJy/beam over 0.9767 GHz. The observations consist of a total of 1920 channels with a total bandwidth of 1.875 GHz. The observing parameters are detailed in Table \ref{tab:table2}.

Data were calibrated and imaged using the Common Astronomy Software Applications (\textbf{CASA}) package \citep{bean2022casa}, pipeline v5.1.1-5, and our own scripts to perform the imaging and post-processing tasks under \textbf{CASA} v5.4.0-70. For the image reconstruction, we used the standard \textbf{hogbom} deconvolution algorithm with \textbf{briggs} weighting and a robustness value of $2.0$. The beam size for our object in the CO(2-1) spectral window is $12.8^{\prime\prime} \, \times \, 6.5^{\prime\prime}$ at a position angle of $PA \,=\, -59.3^{\circ}$, corresponding to $438 \, \times \, 345$ $\rm\,pc^2$ at a distance of $39$ Mpc. The field of view (FOV) has a diameter of 27$^{\prime\prime}$ ($5.1$ kpc) and is covered using a single pointing, with the largest angular scale recoverable being $2.6^{\prime\prime}$ ($\sim \, 490$ pc). The continuum sensitivity of the data is $0.1$ mJy/beam, while the rms noise for the spectral line data, with a channel width of $10$ \, km/s, is $0.7$ mJy/beam.

After reconstructing the image, we prepared the final CO data cube and maps. We used the same deconvolution algorithm for image reconstruction, processing it initially at the native velocity resolution of 2.4 km/s, as shown in Table ~\ref{tab:table2}. To improve the signal-to-noise ratio, we resampled the data to a coarser velocity resolution of 10 km/s. This resampling helped us optimise the analysis of kinematic features while preserving critical details. We applied a masking process to isolate CO(2-1) emission regions, starting with a threshold mask at $2 \times \text{rms}$ noise level to detect significant emission. We then made manual adjustments to refine the mask and minimise noise. This process follows the methodology used in the analysis of ESO 420-G13 \citep{fernandez2020co}. After masking, we subtracted the continuum emission using a zero-degree polynomial fit in the spatial frequency domain and the adjacent channels. Finally, we corrected the data cubes for primary beam attenuation within the 37.84$^{\prime\prime}$ field of view, ensuring accurate calibration for further analysis.

\subsection{Ancillary data}

To gain a more comprehensive understanding of our object, we expanded our analysis beyond ALMA observations. Specifically, we adopted a multi-wavelength approach to construct a well-sampled SED. For this purpose, we utilise Johnson photometry for the U, B, and V bands \citep{mcalary1983near}. Additionally, we incorporated far-ultraviolet (FUV) and near-ultraviolet (NUV) measurements from GALEX \citep{bouquin2018galex}.
\textit{Spitzer}-IRS high-resolution spectra are crucial for constraining the dusty torus component at its peak wavelengths \citep{tommasin2010spitzer}. 2MASS measurements were adopted for the near-infrared continuum in the J and H bands \citep{jarrett20032mass}, and additional 2$\mu$m observations were obtained from \citet{skrutskie2006two}. The Wide-field Infrared Survey Explorer (WISE) survey contribute key data points in the mid-infrared range \citep{cutri2012explanatory}.
HST/STIS (UV, NIR, and optical) observations were collected from the HST science archive (archival research programme ID: HST-AR-16143; PI: M. Malkan) \citet{malkan1998hubble} and shown Fig.~\ref{fig:hst} \citep{cackett2018accretion} together with the CO(2-1) emission.

In order to provide a comprehensive view of possible AGN feeding and feedback mechanisms within host galaxies, our study incorporates SED fitting alongside our CO observations \citep{ciesla2015constraining}. This method enriches our analysis by enabling detailed quantification of dust and gas properties, crucial for interpreting molecular gas dynamics. By linking SED-derived parameters such as dust mass, SFRs, and the AGN's Eddington ratio with CO kinematics, we can better understand the interplay between AGN activity and star formation.  Furthermore, SED fitting allows us to identify the AGN component or fraction and assess its impact on the surrounding environment\citep{atlee2011multi,karouzos2014tale,salome2023star}. We have complemented our observations with additional ALMA data, specifically for the continuum emission in Band 3. This observation was carried out under the project code: 2018.1.00576.S. The selected frequency band for this observation is Band 3, which ranges from 89.509 to 105.494 GHz. The angular resolution achieved during this observation is 0.25$^{\prime\prime}$. The Band 3 observation has a continuum sensitivity of 0.0259 mJy/Beam, obtained with an integration time of 5 minutes. The detailed information of the observations with ALMA is listed in Table~\ref{tab:table2}. The photometry data used for the SED fitting, supported by ancillary data are listed in Table~\ref{table:3}.

\section{Results}\label{res}
  
\subsection{Molecular gas properties}
\begin{figure*}[!htbp]
\centering
    \includegraphics[width=8.5cm]{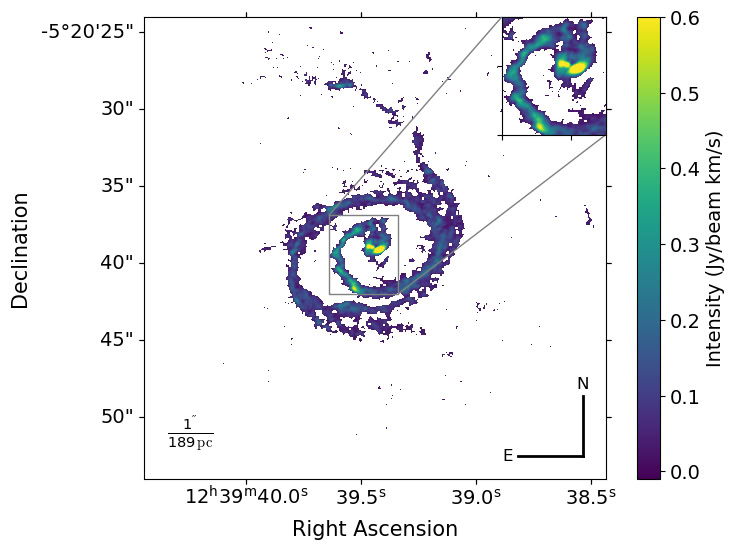}
    \includegraphics[width=8.5cm]{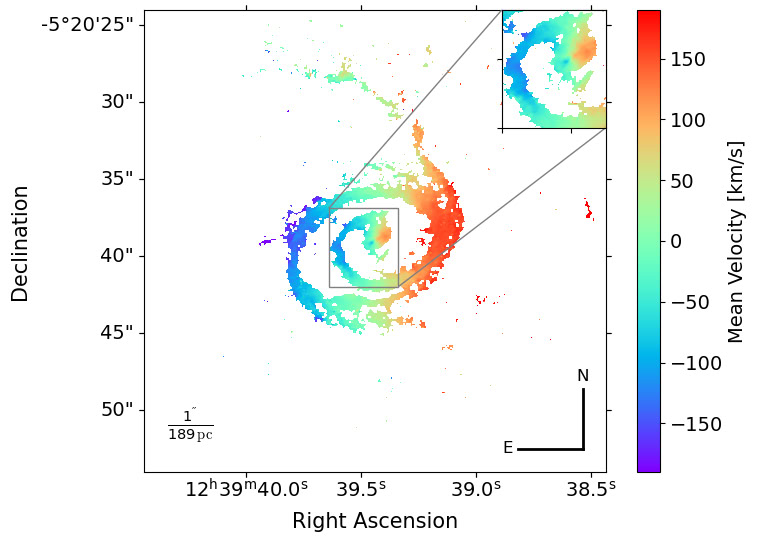}
    \includegraphics[width=8.5cm]{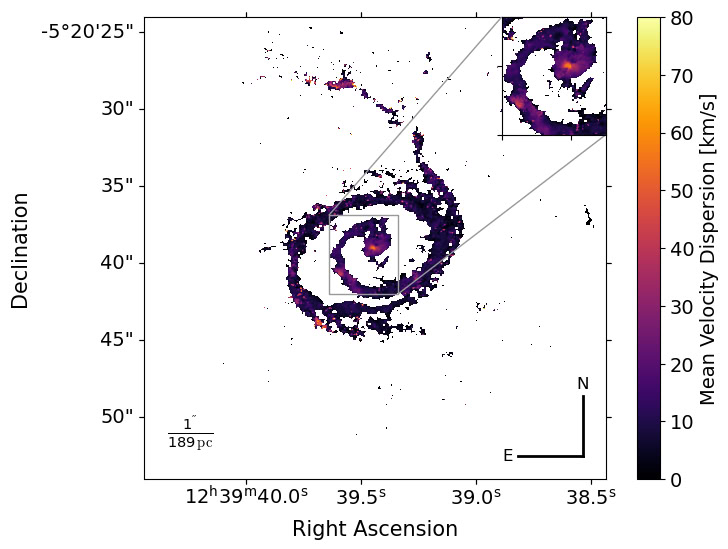}
    \includegraphics[width=8.5cm]{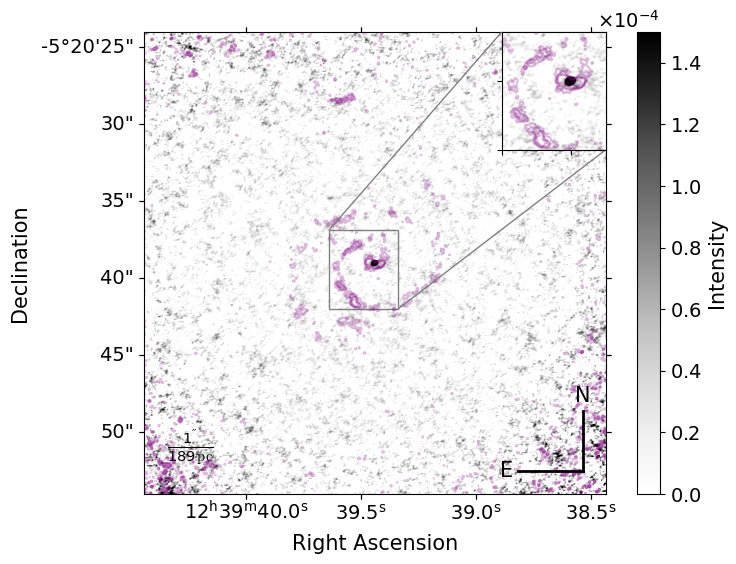}
    \caption{CO(2-1) moment maps from the ALMA Band 6 data cube of NGC\,4593, created using CASA. Top Left: Integrated intensity (moment-0 map) showing the spatial distribution of the line flux, with a total calculated flux of approximately 100 $\rm Jy \, km/s$. The colour bar represents the intensity scale in $\rm Jy \, km/s$ per beam units. Top Right: Intensity-weighted velocity (moment-1 map) with scale bar units in $\rm km/s$, tracing gas velocities and kinematics within the galaxy. Bottom Left: Intensity-weighted velocity dispersion (moment-2 map) with scale bar units in $\rm km/s$. Bottom Right: ALMA Band 6 continuum image (background) displaying the dust distribution, overlaid with CO(2-1) emission moment-0 map contours from ALMA observation Project ID: 2017.1.00236.S.}
    \label{fig:Moments}
\end{figure*}
\begin{table*}[h!]
\centering
\caption{Photometry data for NGC\,4593}
\label{table:3}
\begin{tabular}{l l l l l}
\hline\hline
Observed band & Flux & Significance & Systematic Error (\%) & Ref. \\
\hline
\textit{Swift} ($2$--$10$ keV)  & $3.16 \times 10^{-11} \, \mathrm{erg\,cm^{-2}\,s^{-1}}$ & No uncertainty reported & 10\% & (1)\\
\textit{Swift} ($14$--$195$ keV) & $3.4 \times 10^{-10} \, \mathrm{erg\,cm^{-2}\,s^{-1}}$ & 90\% confidence & 10\% & (1)\\
GALEX (FUV) & $3.60 \pm 0.05 \, \mathrm{mJy}$ & $1\sigma$ & 5\% & (2)\\
GALEX (NUV) & $5.91 \pm 0.03 \, \mathrm{mJy}$ & $1\sigma$ & 5\% & (2)\\
2MASS (J) & $415.00 \pm 7.71 \, \mathrm{mJy}$ & $1\sigma$ & 2\% & (3)\\
2MASS (H) & $494.00 \pm 10.20 \, \mathrm{mJy}$ & $1\sigma$ & 2\% & (3)\\
2MASS (Ks) & $427.00 \pm 12.40 \, \mathrm{mJy}$ & $1\sigma$ & 2\% & (4)\\
IRAC ($3.6\,\mu\mathrm{m}$) & $125.00 \pm 8.03 \, \mathrm{mJy}$ & $1\sigma$ & 5\% & (5)\\
IRAC ($4.5\,\mu\mathrm{m}$) & $134.00 \pm 21.00 \, \mathrm{mJy}$ & $1\sigma$ & 5\% & (5)\\
IRAC ($5.8\,\mu\mathrm{m}$) & $163.00 \pm 23.95 \, \mathrm{mJy}$ & $1\sigma$ & 5\% & (5)\\
IRAC ($8.0\,\mu\mathrm{m}$) & $229.00 \pm 33.16 \, \mathrm{mJy}$ & $1\sigma$ & 5\% & (6)\\
IRAS ($12\,\mu\mathrm{m}$) & $299.00 \pm 73.00 \, \mathrm{mJy}$ & Uncertainty, confidence not specified & 10\% & (7)\\
Spitzer MIPS ($24\,\mu\mathrm{m}$) & $633.00 \pm 89.00 \, \mathrm{mJy}$ & No uncertainty reported & 10\% & (5)\\
ALMA (Band 6) & $8.13 \pm 0.22 \, \mathrm{mJy}$ & $1\sigma$ & 10\% & (8)\\
ALMA (Band 3) & $8.42 \pm 0.06 \, \mathrm{mJy}$ & $1\sigma$ & 10\% & (8)\\
\hline
\end{tabular}
\tablefoot{
This table presents the photometry data for NGC\,4593 across a range of observed bands, including X-ray, UV, near-infrared, and mid-infrared wavelengths. Flux values are given in units of $\mathrm{mJy}$, except for the \textit{Swift} X-ray data, which are expressed in $\mathrm{erg\,cm^{-2}\,s^{-1}}$. The uncertainties listed in the flux column correspond to the statistical errors of the measurements. The significance column denotes the confidence level associated with the data, while the systematic error indicates the potential percentage uncertainty due to instrumental calibration or observational conditions.
}

\tablebib{
(1)~\citet{ricci2017bat}; (2)~\citet{bouquin2018galex}; (3)~\citet{mcalary1983near}; (4)~\citet{jarrett20032mass}; (5)~\citet{skrutskie2006two}; (6)~\citet{cutri2012explanatory}; (7)~\citet{tommasin2010spitzer}; (8)~Obtained in this work based on ALMA observation project IDs: 2017.1.00236.S and 2018.1.00576.S.
}
\end{table*}

Moment maps were created from the available data cubes and are shown in Fig.~\ref{fig:Moments}. The moment-0 map delineates the overall spatial molecular gas distribution within the galaxy, indicating intensity values ranging from 0 to 6.0 $\rm Jy/beam \cdot km/s$, with a heightened concentration of gas discernible in the very inner regions of the galaxy. The moment-1 map traces the velocity field of the gas, quantified in units of $\rm km/s$, and highlights the dynamics in NGC\,4593, including the rotational dynamics and the kinematic patterns inherent in the galaxy's structure. In Fig.~\ref{fig:Moments}, we present a comparison between the dust and molecular gas distributions within the galaxy under study. This figure features the continuum emission, indicating the extent of dust, with ALMA CO(2-1) contours overlaid to highlight the molecular gas distribution.

As shown in Fig.~\ref{fig:Moments}, the CO gas distribution within NGC\,4593 exhibits a striking one-arm structure. This feature is indicative of a logarithmic $m=1$ mode, which refers to a pattern where there is a single dominant spiral arm in the distribution of gas. This phenomenon is not commonly observed in the nuclear regions of barred galaxies \citep{shaw1995nuclear, phookun1993ngc, ann2005formation, thakur2009effect}. The $m=1$ mode suggests a significant deviation from axisymmetric equilibrium, potentially driven by tidal interactions or internal instabilities.

\begin{figure*}[!ht]
\centering
    \includegraphics[width=17cm]{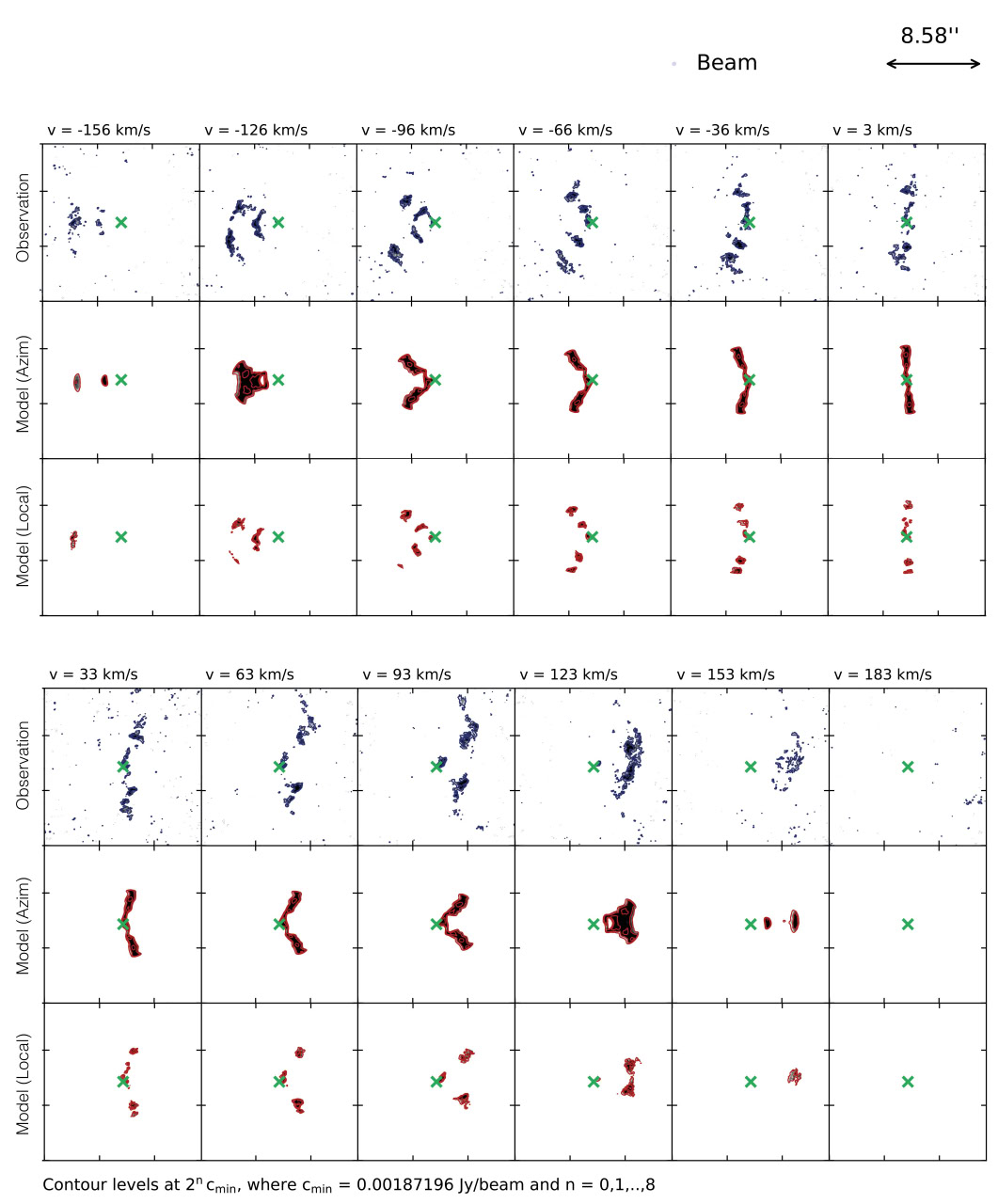}
    \caption{CO(2-1) channel maps from the ALMA B6 data cube, generated using \textsc{3D-Barolo} modelling, provide insight into the morphology and gas kinematics of NGC\,4593. The channel maps display velocities ranging from $-230$ $\rm km/s$ (top left) to $+156$ $\rm km/s$ (bottom right), relative to the galaxy's systemic velocity ($v_{\text{sys}}$). This $v_{\text{sys}}$ is estimated as the central velocity in the global line profile, valued at $12.56$ km/s. The velocities presented signify deviations from this systemic velocity. Maps are shown in increments of $10$ $\rm km/s$, and each square box covers an area of 8.58$^{\prime \prime}$. The top row displays selected channels in blue, while the middle and bottom rows depict maps normalised using the azimuthal and local normalisation methods, respectively, highlighting distinct patterns in the emission's spatial distribution and intensity. A green cross in each panel marks the disc's centre, and solid red lines represent the RMS noise level ($\sigma_{\text{rms}}$) of $0.7$ $mJy$.}
    \label{fig:chan_azim}
\end{figure*}

\subsection{\textsc{3D-Barolo} model}

To delve into the kinematics of NGC\,4593's innermost regions, we employed the 3D-Based Analysis of Rotating Object via Line Observations (\textsc{3D-Barolo}) software \citep{di20153d}. This tool fits a 3D tilted-ring model to emission line data cubes, enabling us to infer the galaxy's rotation curves and the gas column density, a critical parameter in our analysis. \textsc{3D-Barolo} provides two distinct normalisation options for the gas column density: pixel-by-pixel (\textbf{LOCAL}) and azimuthally averaged (\textbf{AZIM}). The \textbf{LOCAL} normalisation allows to consider the non-axisymmetric density models and prevents regions with unusual gas distributions, such as clumpy regions or voids, from skewing the overall fit. This method offers a granular view of the galaxy's kinematics and is particularly beneficial when examining complex structures or regions with high spatial variability in the gas distribution. On the other hand, the \textbf{AZIM} normalisation uses the azimuthally averaged flux in each ring as the normalisation value, providing a more global perspective of the galaxy's kinematics. This method is particularly effective when the goal is to understand large-scale kinematic patterns in the galaxy, such as rotation curves or velocity dispersion profiles. 

\begin{figure*}[!ht]
\centering
    \includegraphics[width=8.5cm]{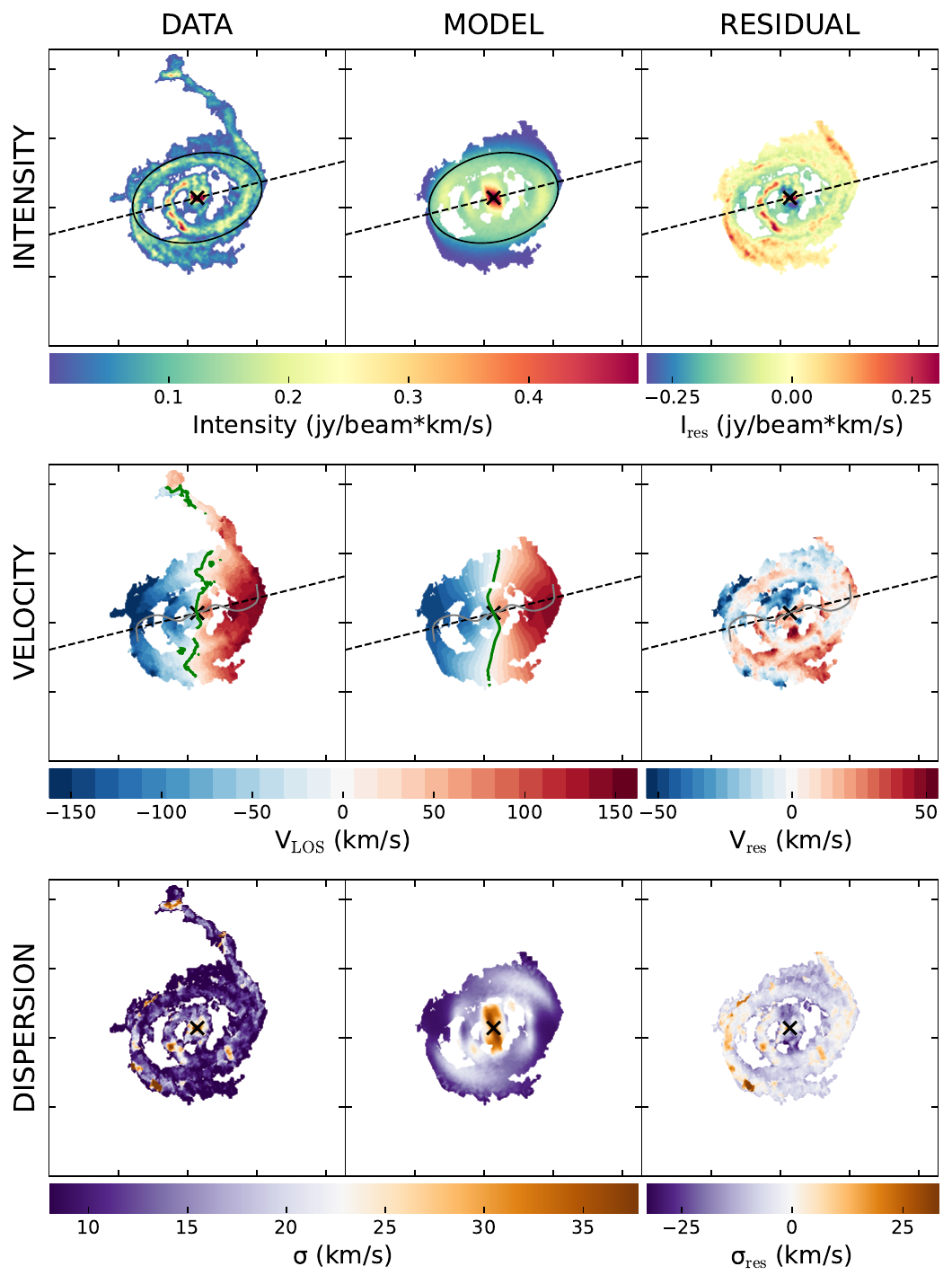}
    \includegraphics[width=8.5cm]{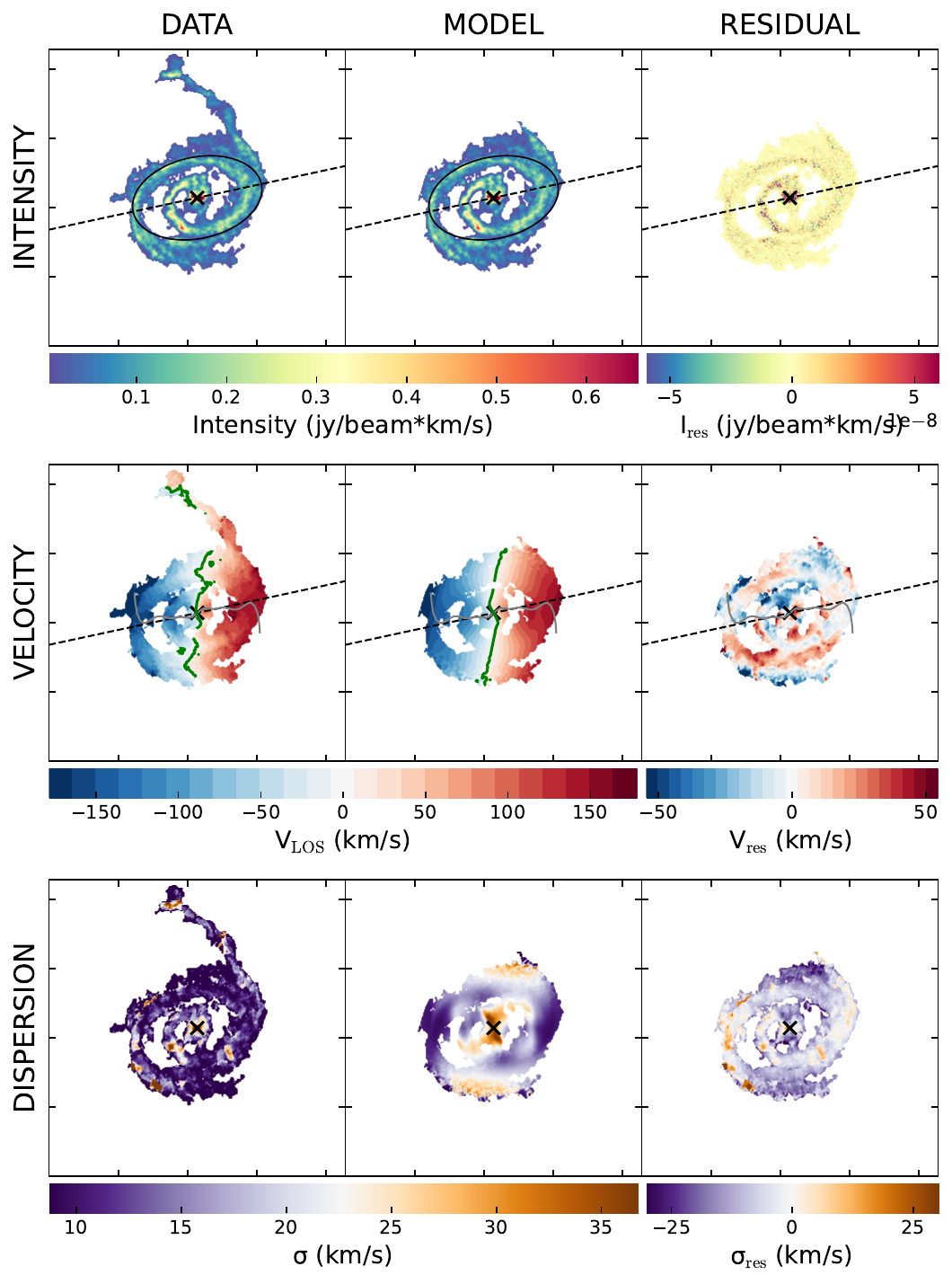}
    \caption{Detailed intensity, velocity, and dispersion maps for the CO(2-1) emission in NGC\,4593, developed using \textsc{3D-Barolo} software. The figure is divided into two panels: the left panel refers to the azimuthal normalised model, while the right one refers to the local normalised model. Each panel consists of a 3x3 grid of images (rows: intensity, velocity, dispersion; columns: data, model, residuals). The top row presents the CO(2-1) intensity maps across the field of view, with the colour bar representing the intensity scale in $Jy \, km \,s^{-1}$ per beam units. The middle row shows the velocity maps, depicting the gas velocities across the field of view, with scale bar units in $\rm km/s$. The bottom row displays the velocity dispersion of the CO(2-1) line across the field of view, also denoted in $\rm km/s$. The left column in each panel shows the data, the middle column displays the model, and the right column provides the residuals.}
    \label{fig:maps}
\end{figure*}

We used the CO(2-1) data cube with a spectral resolution of 10 $\rm km/s$ as input data for \textsc{3D-Barolo}. The tilted-ring model adopted by \textsc{3D-Barolo} operates on two fundamental assumptions: the emitting material is encapsulated within a thin disc, and the angular momentum of the galaxy is responsible for the observed rotational pattern. These assumptions prove to be remarkably effective in scrutinising galaxy kinematics, extending their applicability even to scenarios where the galaxy is not perfectly edge-on or the gas does not adhere to a circular motion trajectory. As inputs to the model, we used \textit{Hubble} and Pan-STARRS images of the galaxy to deduce the initial values for position angle (PA) and inclination (INC). The ALMA moment-0 maps (see Fig.~\ref{fig:Moments}) are instrumental in pinpointing the galaxy's centre and systemic velocity ($\rm V_{SYS}$). Fig.~\ref{fig:chan_azim} showcases the channel maps, which feature two distinctive normalisations, \textbf{LOCAL} and \textbf{AZIM}, conceived through the analytical process of \textsc{3D-Barolo}.

Fig.~\ref{fig:maps} shows a dissection of the molecular gas dynamics in NGC\,4593, the CO(2-1) emission characteristics within the galaxy, as interpreted through two distinct normalisation methods: Azimuthal and Local, respectively portrayed in separate panels. The model enables us to discern a clear velocity field within the residual representations, indicating a spread of up to $\pm \, 50 $ $\rm km/s$. This trend is consistent across both the Azimuthal and Local normalisation techniques, substantiating the robustness of the observed patterns. Similarly, the velocity dispersion maps delineate variations up to $50$ $\rm km/s$, shedding light on the complex, yet orderly gas dynamics prevalent within the NGC\,4593 galaxy. Furthermore, the velocity field distribution ranges from $-150$ to $150 \, \rm km/s$ for both data and model, while the residuals show a range of -50 to 50 $\rm km/s$. The dispersion (sigma) ranges from $10$ to $35$  $\rm km/s$ for data and model, and $-10$ to $20$ $\rm km/s$ for residuals. The intensity ranges from $0.002$ to $0.016$ $\rm Jy \, km/s$ for data/model and $-0.0025$ to $0.0025$ for residuals.

\begin{figure}[!htbp]
\resizebox{\hsize}{!}{   \includegraphics{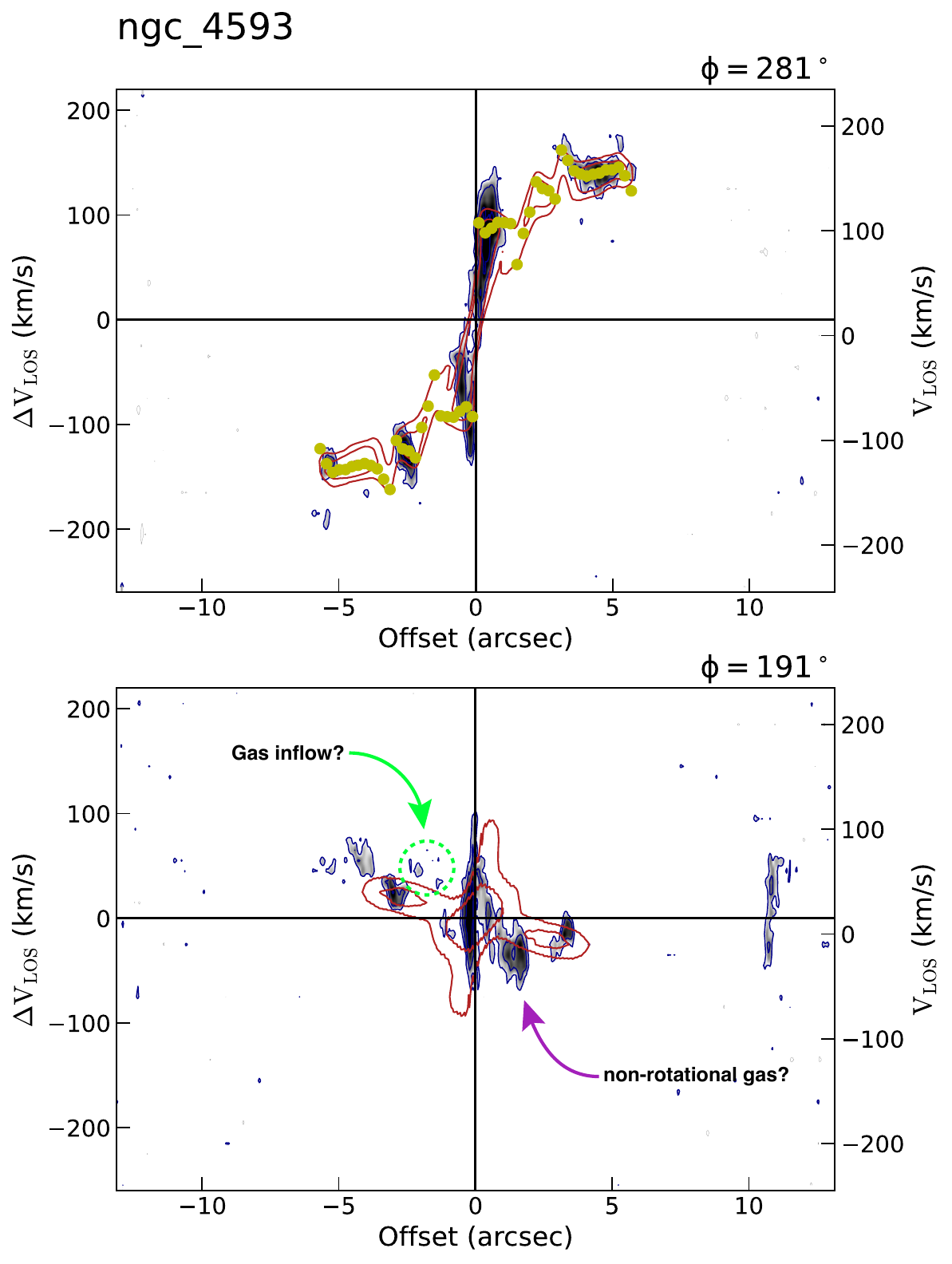}}
    \caption{Position-velocity (p-v) diagrams of the CO(2-1) emission in NGC\,4593, extracted from the data and \textsc{3d-Barolo} model along the major (top panel) and minor (bottom panel) axes. Blue solid contours illustrate the p-v diagram derived from the data cube, while red solid contours depict the p-v diagram from the model cube. The contour levels for both data and model are set at $[1, 2, 4, 8, 16, 32, 64] \, \times\, l$, with $l = 0.0012$. In the top panel, yellow solid dots represent the rotation velocity of each ring from the best-fit disc model. The dashed light green circle in the bottom panel highlights the region where a potential gas inflow or outflow is located. These p-v diagrams facilitate the investigation of the spatial and kinematic structure of the CO(2-1) emission in NGC\,4593.}
    \label{fig:pv}
\end{figure}

The position-velocity ($p-v$) diagrams of the CO(2-1) emission in NGC\,4593, extracted from the data and model cubes along the major and minor axes, are shown in Fig.~\ref{fig:pv}. For the major axis (top panel), the observed rotational velocities (yellow dots) from the best-fit disc model align closely with the data-derived contours, showcasing a consistent rotation curve. The minor axis (bottom panel), on the other hand, offers insights into non-rotational gas motions 1.5$^{\prime\prime}$ away from the centre of the galaxy (redshifted motion to the south of AGN near the centre). The dashed green circle in the bottom panel highlights a region about 2$^{\prime\prime}$ away from the centre, where possible gas inflow or outflow is observed. The disparities between the model and observed data in this plane may hint at vertical motions or inflows (or outflows), potentially driven by mechanisms like feeding or feedback, interactions with neighbouring galaxies, or even AGN-driven winds. In the bottom panel of Fig.~\ref{fig:pv}, it is clear that the model does not accurately represent the molecular gas distribution between 4$^{\prime\prime}$ and 6$^{\prime\prime}$. This discrepancy could arise from the model primarily emphasising the central region. Alternatively, it might indicate the presence of cold molecular outflows.

\subsection{\textsc{discFit} model}

\begin{figure*}[!htbp]
\centering
    \includegraphics[width=17cm]{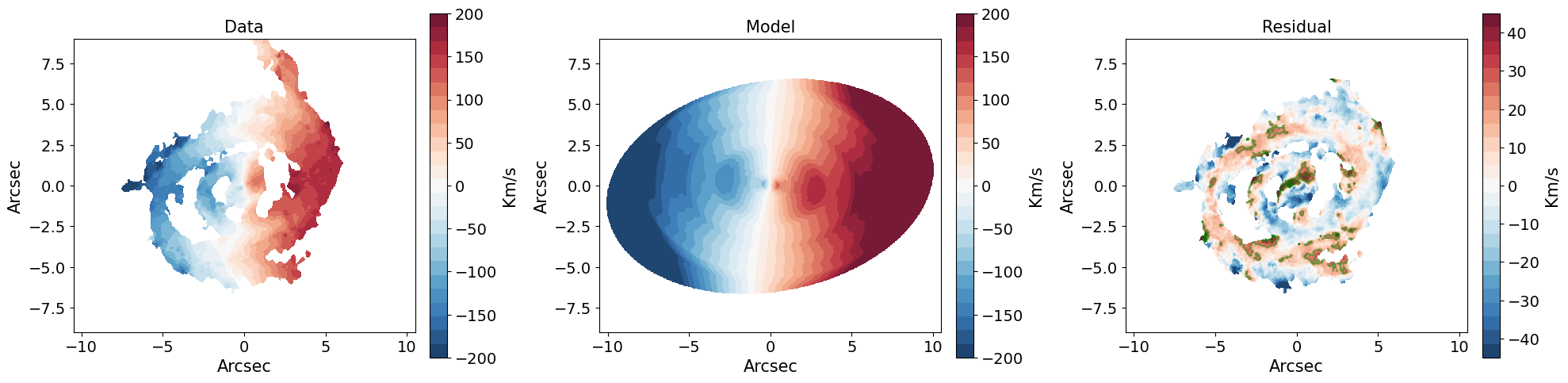}
    \includegraphics[width=12cm]{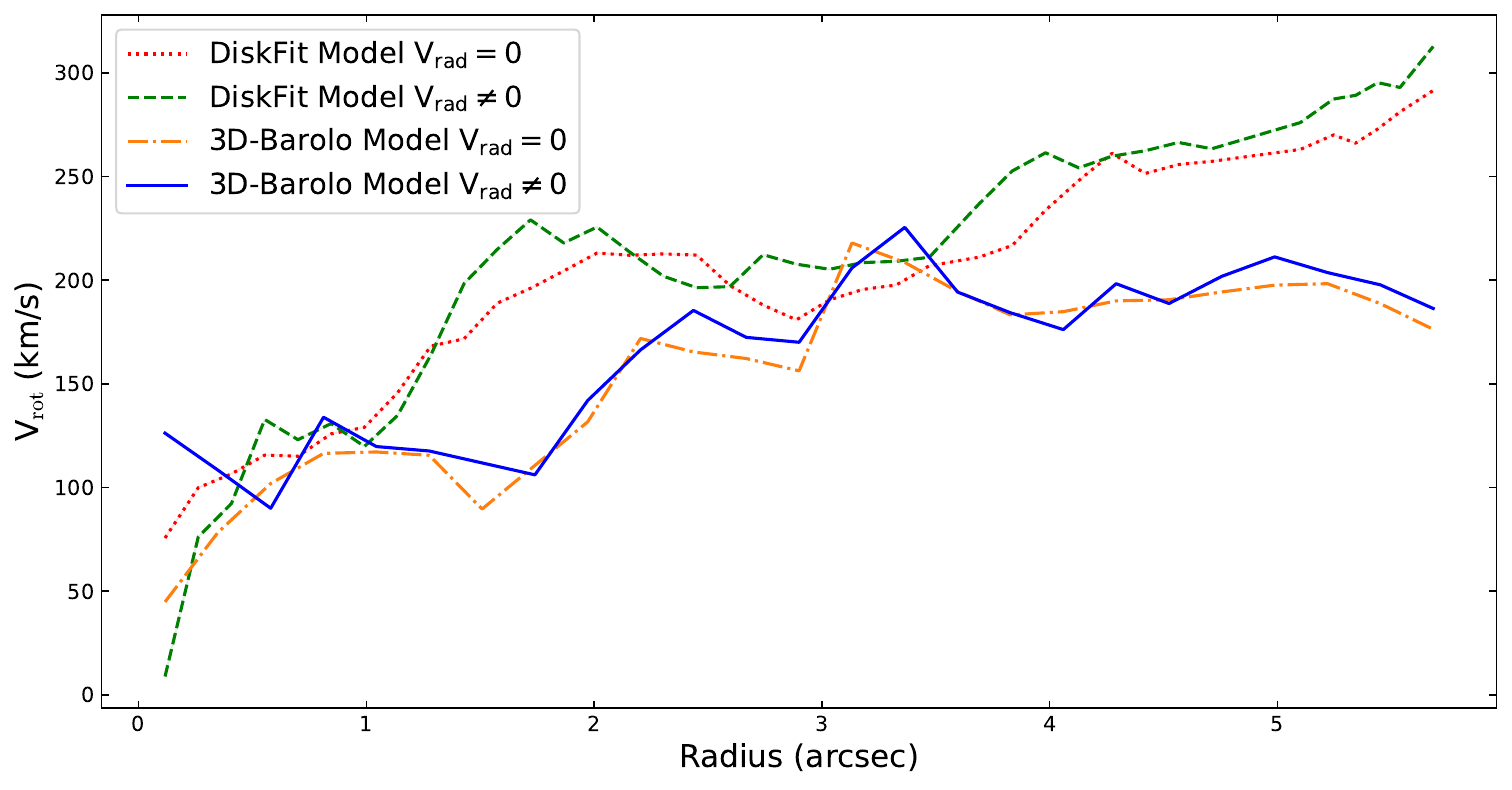}
    \caption{CO(2-1) observed velocity field and corresponding residuals derived from \textsc{discFit} modelling for NGC\,4593. The left panel showcases the observed velocity field of the CO(2-1) emission, with the scale bar units given in $\rm km/s$. The middle panel shows the residuals produced by subtracting the \textsc{discFit} model from the observed data, with green contours representing levels of $5\sigma_{rms}$. The bottom panel compares the rotation curve velocities ($V_{rot}$) obtained from both \textsc{3D-Barolo} and \textsc{discFit} models, considering cases with and without radial velocity ($V_{rad}$).}
    \label{fig:discfit}
\end{figure*}

To complement and cross-verify our results obtained from the \textsc{3D-Barolo} analysis, we utilised an additional analysis tool, \textsc{discFit}. \textsc{discFit} \citep{spekkens2007modeling} is an instrumental modelling tool for dissecting the structural components of galaxies. This tool adeptly deconstructs a galaxy's luminosity distribution into its fundamental constituents, such as the disc, bulge, and bar structures, when they are present. We applied \textsc{discFit} to NGC\,4593 and replicated the galaxy's observed kinematic features using the velocity moment map as an input, and utilising the same initial conditions as used for \textsc{3D-Barolo}. However, as shown in Fig.~\ref{fig:discfit} (residuals in the top right and rotation curve comparison in the bottom plot), we observe notable deviations, particularly in the galaxy's core and along the first eastern spiral arm. These deviations, visible in the residuals and the rotation curve comparison, show that while the models capture the general kinematic structure, they struggle to accurately represent the complex dynamics in these regions. One key observation is that \textsc{3D-Barolo} tends to align more closely with observational data when accounting for radial velocity, suggesting a potential overfitting to axisymmetric non-circular motions. In contrast, \textsc{discFit} provides a more consistent fit regardless of radial velocity inclusion. Notably, \textsc{discFit} does not significantly react to data bumps that \textsc{3D-Barolo} attempts to follow. A specific example of this behaviour is evident near the galaxy's centre, around 0.8$^{\prime\prime}$ from the nucleus, which can be observed in the residual image of \textsc{discFit} overlaid with contours.

The discrepancy between models could also be attributed to their dimensional approaches: \textsc{discFit}'s 2D modelling contrasts with \textsc{3D-Barolo}'s additional consideration of disc scale-height (Z0 parameter), set to 1$^{\prime\prime}$ in our study. This difference becomes particularly apparent at the edges of the rings in the models. \textsc{3D-Barolo} excels in detailing galaxy morphology, aiding in the derivation of physical quantities like surface density and the calculation of molecular and black hole masses. Conversely, \textsc{discFit} better elucidates axisymmetric behaviours. However, the face-on orientation of NGC\,4593 complicates the evaluation of model accuracy. Furthermore, the absence of the galaxy's outer northern arms, approximately 10$^{\prime\prime}$ from the centre, raises questions about possible outer rings or gas stripped and scattered in a conical distribution.

\begin{figure}[!htbp]
\resizebox{\hsize}{!}{
    \includegraphics{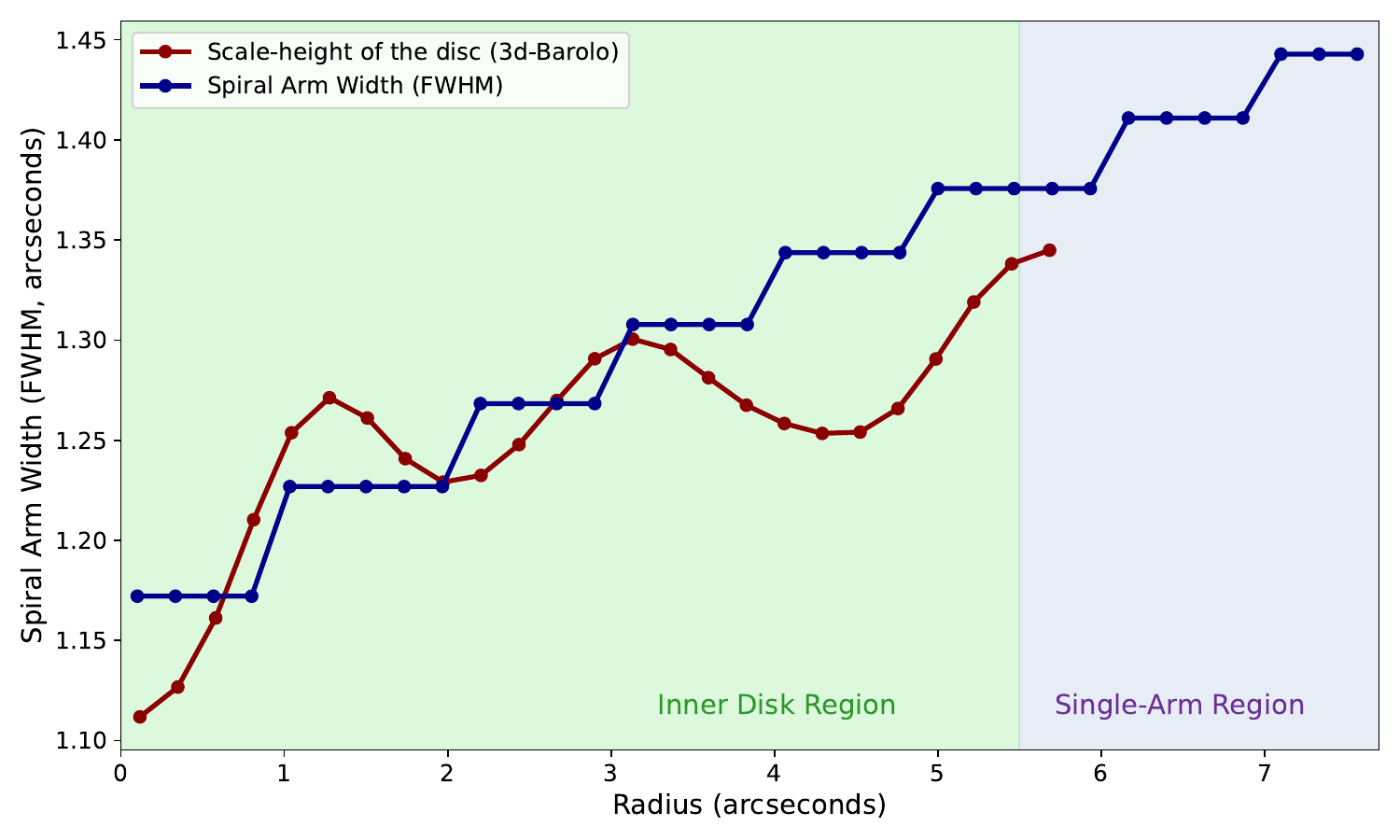}}
    \caption{Comparison between the spiral arm widths and disc scale height as a function of radius in NGC\,4593. The blue line represents the spiral arm widths, estimated from the observational data of CO(2-1) molecular gas. The red line shows the disc scale height, derived from \textsc{3D-Barolo} modelling.}
    \label{fig:spiral_arm_widths}
\end{figure}

In Fig.~\ref{fig:spiral_arm_widths}, we compare the spiral arm widths, estimated from the observational data of CO(2-1) molecular gas, with the disc scale height derived from \textsc{3D-Barolo} modelling. In the inner disc region (0 to 5.5$^{\prime\prime}$), the spiral arm width ranges from 1.0$^{\prime\prime}$ to 1.3$^{\prime\prime}$, while the disc scale height shows a similar variation, from 1.1$^{\prime\prime}$ to 1.3$^{\prime\prime}$. Beyond 5.5$^{\prime\prime}$, the single-arm region has a width ranging from 1.35$^{\prime\prime}$ to 1.45$^{\prime\prime}$. Although it is not possible to measure the flaring directly without an edge-on galaxy, the proportionality between the width and height suggests that the flaring in NGC\,4593 is compatible with that seen in other galaxies \citep[e.g.,][]{patra2020theoretical}.

\section{Discussion}\label{Discussion}

\subsection{Gas kinematics}

\begin{figure*}[!htbp]
\centering
    \includegraphics[width=17cm]{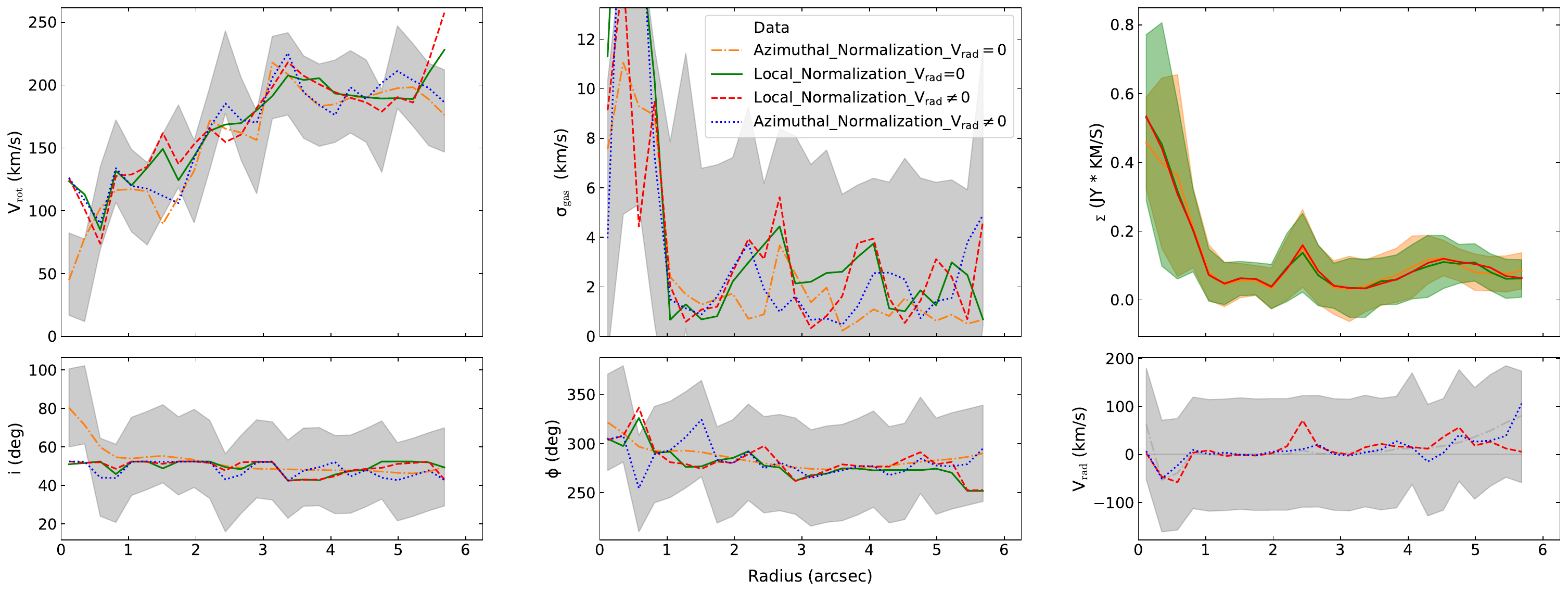}
    \caption{Kinematic and structural parameters of NGC\,4593 as a function of radius. Top row, from left to right: \textbf{Rotation Velocity ($V_{\mathrm{rot}}$)}, \textbf{Velocity Dispersion ($\sigma_{\mathrm{gas}}$)}, and \textbf{Surface Density ($\Sigma$)}. Bottom row, from left to right: \textbf{Inclination Angle ($i$)} showing the variation of inclination with radius, \textbf{Position Angle ($\phi$)} illustrating the fluctuation of position angle with radius, and \textbf{Radial Velocity ($V_{\mathrm{rad}}$)} displaying the radial velocity ($\rm km/s$) versus radius, marked in grey. In each subplot, the observational data are represented by grey points with error bars. The fitted azimuthal normalisation with $V_{\mathrm{rad}}=0$ is depicted by an orange line with a dash-dot pattern, while the local normalisation series with $V_{\mathrm{rad}}=0$ is shown as a solid green line. The local normalisation with $V_{\mathrm{rad}}\neq 0$ is represented by a dashed red line, and the azimuthal normalisation with $V_{\mathrm{rad}}\neq 0$ is indicated by a dotted blue line.}
    \label{fig:parameters}
\end{figure*}

In our analysis, delineated in Fig.~\ref{fig:parameters}, we subjected the ALMA data for NGC\,4593 to a detailed \textsc{3D-Barolo} analysis. This analytical representation comprises six plots, each detailing a distinctive parameter — $V_{\text{rot}}$ (rotation velocity), $\sigma_{\text{gas}}$ (gas velocity dispersion), $v_{\text{rad}}$ (radial velocity), $\phi$ (azimuthal angle), $i$ (inclination angle), and $\Sigma$ (gas surface density measured in $\rm Jy \, km/s$) — in relation to the radius measured in arcseconds.

The best-fit values of the model to the data, derived from this analysis, indicate an average inclination of 47.4 degrees and a position angle (PA) of 276 degrees. Since the kinematic major axis is roughly E-W, the near side is located to the north, and the far side to the south, as indicated by the molecular gas kinematics (see Fig.~\ref{fig:velocityfield}). We conclude that the near side of NGC\,4593 is to the north and the far side to the south, further supported by the dust lane morphology visible in HST images (see Fig.~\ref{fig:hst}). This conclusion is confirmed by extended MUSE observations of the galaxy's kinematics \citep{den2020muse}, which ensure consistency with the northern side as the near side.

The residual map reveals deviations from the expected rotation (see Figs.~\ref{fig:maps},\,~\ref{fig:discfit}). A redshifted residual does not necessarily imply that a region is receding; it may instead indicate less blueshifted velocity than expected. These residuals could result from unmodelled radial gas motions or other dynamic factors. \textsc{3D-Barolo}'s axisymmetric assumption may not fully account for the kinematics of molecular gas in heavily barred systems. For instance, \citet{schnorr2014feeding} demonstrated that non-axisymmetric bar potentials in NGC2110 influenced significant molecular outflows, suggesting that similar effects could be at play in NGC\,4593.

\subsection{Estimation of the molecular gas mass}

We computed the CO(2–1) line luminosity \( L^{\prime}_{\text{CO}(2-1)} \) using the following equation \citep{solomon2005molecular}:
\begin{equation}
L^{\prime}_{\text{CO}(2-1)} = 3.25 \times 10^{7} \, S_{\text{CO}}\Delta v \, \nu^{-2}_{\text{obs}} \, D^{2}_{L} \, (1+z)^{-3}.
\end{equation}
To determine the total flux \( S_{\text{CO}}\Delta v \), we used the moment-0 map of the ALMA CO(2-1) observations (Fig.~\ref{fig:Moments}), where the restoring beam size is \( 0.23^{\prime \prime} \times 0.18^{\prime \prime} \) at a position angle of \(-59.30\) degrees. Focusing on a region with a \( 12^{\prime \prime} \) radius, we obtained a flux density of \( 146.3 \pm 1.1 \) $\rm Jy \,\rm km/s$, which serves as our value for \( S_{\text{CO}}\Delta v \). Using these values, along with \( D = 39 \) Mpc, \( z = 0.009 \), and \( \nu_{\text{obs}} = 230 \) GHz, we calculated the CO(2–1) line luminosity to be:
\begin{equation}
L^{\prime}_{\text{CO}(2-1)}  \approx 1.40 \pm 0.10\times 10^{8} \,\, [\text{K km s}^{-1} \text{ pc}^{2}].
\end{equation}

We then converted \( L^{\prime}_{\text{CO}(2-1)} \) to \( L^{\prime}_{\text{CO}(1-0)} \) using a conversion ratio of \( \frac{L^{\prime}_{\text{CO}(2-1)}}{L^{\prime}_{\text{CO}(1-0)}} \approx 5.0 \), which reflects the typical ratio observed in galaxies with moderate to high levels of star formation, including (ultra) luminous infrared galaxies ((U)LIRGs) \citep{kamenetzky2016relations}. This conversion yields:
\begin{equation}
L^{\prime}_{\text{CO}(1-0)} \approx (0.28 \pm 0.02) \times 10^{8} \,\, [ \text{K km s}^{-1} \text{ pc}^{2}].
\end{equation}
Using the conversion factor \( \alpha_{\text{CO}} \), which ranges from \( 0.8 \, M_{\astrosun} \) to \( 3.2 \, M_{\astrosun} \) \citep{bolatto2013co}, we estimated the molecular gas mass \( M(H_2) \) to be within the range:
\begin{equation}
M_{H_2} = \alpha_{\text{CO}} \, L^{\prime}_{\text{CO}(1-0)} \approx (0.22 - 0.90) \times 10^{8} \, M_{\astrosun}.
\end{equation}
It is important to note that, in addition to the conversion factor, there are uncertainties in the derived total CO luminosity due to the nature of interferometric observations, which can filter out extended emissions. Consequently, the total molecular gas mass could be larger if CO(2-1) emission is more extended.

\subsection{Estimation of SMBH mass}

To estimate the mass of the supermassive black hole (SMBH) in NGC\,4593, we adopted the methodology outlined in \citet{smith2021wisdom}, which relies on the correlation between the SMBH mass and the rotation of its host galaxy using molecular gas kinematics. The relationship is described by the following equation:
\begin{equation}
\log \left(\frac{M_{\text{BH}}}{M_{\odot}}\right) = (7.5 \pm 0.1) + (8.5 \pm 0.9) \left[\log \left(\frac{W_{50}}{\sin i} \, \text{km s}^{-1}\right) - 2.7\right],
\end{equation}
where \( M_{\text{BH}} \) represents the central SMBH mass, \( W_{50} \) denotes the full width at half-maximum of a double-peaked CO(2-1) emission line profile, and \( i \) is the inclination of the CO disc. This relationship exhibits a total scatter of \( 0.6 \) dex, comparable to other SMBH mass correlations, which have an intrinsic scatter of \( 0.5 \) dex.

\begin{figure}
\resizebox{\hsize}{!}{   \includegraphics{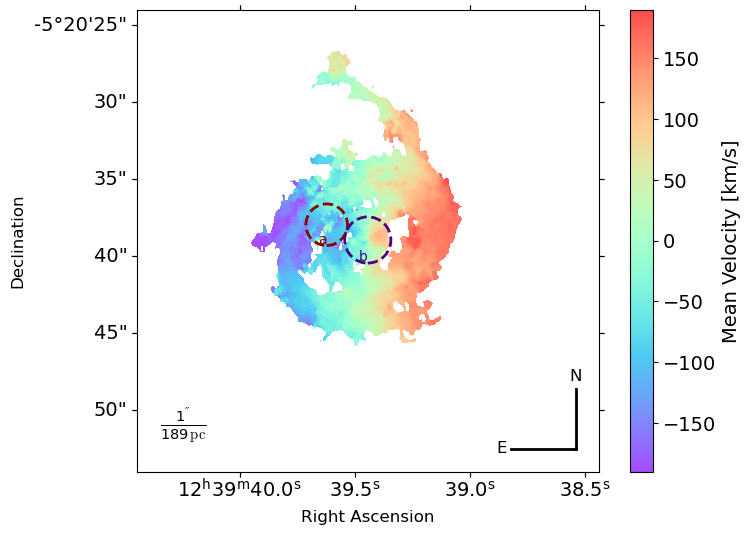}}
   \resizebox{\hsize}{!}{   \includegraphics{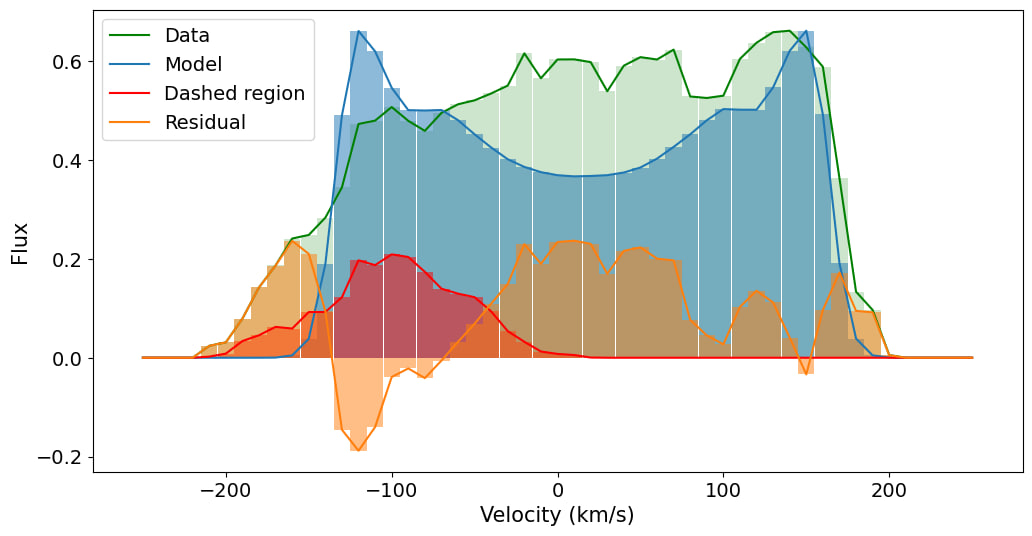} }   
    \caption{Analysis of molecular gas dynamics in NGC\,4593. Top: Red-circled region (a) indicates potential non-circular motion due to the outflow of cold molecular gas, visible on the mean velocity map (background). The central dashed circle (region b) highlights the CMZ-like ring, which contains significant molecular gas. Bottom: Comparison between the CO(2-1) ALMA observations (green) and the \textsc{3D-Barolo} model (blue) for NGC\,4593. The plot shows the average flux as a function of velocity, with residuals (orange) and possible outflowing gas from region (a) highlighted in red.}
    \label{fig:velocityfield}
\end{figure}

For our analysis, we derived the line width at half-maximum, \(W_{50}\), by employing the method described by \citet{tiley2016tully} for measuring the line width of double-peaked profiles. This approach involves fitting a Gaussian double-peak profile to the emission-line data. As demonstrated in Fig.~\ref{fig:velocityfield}, we utilised a double-peak Gaussian function to accurately fit the integrated CO(2-1) emission-line spectrum, representing the observed emission profile from the entire field of view (FOV) as obtained from our \textsc{3D-Barolo} modelling. The $W_{50}$ value was determined using the formula $W_{50} = 2(w + \sigma \sqrt{2 \ln 2})$, where $w$ is the velocity half-width at half-maximum (HWHM) and $\sigma$ is the velocity width of the Gaussian peaks. From this analysis, we obtained $W_{50} \simeq 314 \pm 10$ $\rm km/s$, and the inclination was derived as $i \simeq 47.70^{\circ} \pm 1.1$, as depicted in Fig.~\ref{fig:parameters}.

Combining these measurements, we calculated the SMBH mass as $\log \left( \frac{M_{\text{BH}}}{M_{\odot}} \right) = 6.89 \pm 0.04$. This result was compared with the values obtained from the WISDOM project for NGC\,4593 \citep{tiley2016tully}, yielding $\log \left( \frac{M_{\text{BH}}}{M_{\odot}} \right) = 6.86$, and with \citet{denney2006mass}, who derived an SMBH mass of approximately $\log \left( \frac{M_{\text{BH}}}{M_{\odot}} \right) = 6.99$ using reverberation mapping.
\subsection{Evidence of non-circular motion and molecular inflow or outflows}

Our analysis using \textsc{3D-Barolo} on NGC\,4593 reveals significant dynamics within the galaxy's gas structure. The CO gas distribution exhibits a one-arm structure, indicative of an \( m = 1 \) density wave or perturbation. Such a feature, rare in the nuclear regions of barred galaxies, suggests a dominant regular velocity field, implying limited non-circular motion. However, the presence of this \( m = 1 \) feature challenges the axisymmetric assumptions inherent in the \textsc{3D-Barolo} models. The discrepancy in the fit between the galaxy's northern and southern sections highlights the limitations of current models in fully capturing the kinematic complexity. This aspect aligns with observations by \citet{barbosa2006gemini} regarding ionised gas kinematics, offering a broader comparative perspective on gas dynamics in barred galaxies.

The observed CO ring (inner) near the centre of the bar shows a complex interplay between bar dynamics and molecular gas distribution. The kinematic position angle (PA) of approximately $ -59.3^{\circ}$ in the nuclear disc significantly differs from the morphological PA of $46^{\circ}$, suggesting that the ring has an elliptical structure with elongated stellar and gas orbits. This discrepancy likely indicates non-circular motions within the nuclear region, possibly driven by the influence of the central bar or other asymmetries in the gravitational potential. The inner structure of NGC\,4593 is intricately shaped by its AGN, which exerts considerable influence on both the light profile and the dynamics of the central region, as demonstrated by \citet{alexei2008image}. The poor fit in the nuclear region may reflect the limitations of the current kinematic model, which assumes axisymmetric conditions that may not fully capture the complexities of the actual orbital dynamics.

Adding to this complexity, the misalignment between its bar and bulge, alongside the presence of a pseudobulge identified through Ks-band surface photometry \citep{kormendy2006pseudobulges}, suggests a deviation from the standard evolutionary path driven by secular processes and bar-driven gas inflows \citep{sanders2022mira}. Furthermore, the central region is characterised by a nuclear dust ring linked to radial dust lanes, indicating active gas inflow and potential starburst activity, which could significantly impact the pseudobulge's development. The dynamic evolution of this region is further influenced by a non-axisymmetric bar potential, which alters gas dynamics and morphological structures, marking a critical phase in the galaxy's evolution.

Residuals from the \textsc{3D-Barolo} modelling on velocity dispersion maps (Fig.~\ref{fig:maps}) further support our findings by highlighting regions of excess velocity dispersion. These residuals, unaccounted for by the model, suggest additional kinematic components like turbulence or unresolved inflows (or outflows). Fig.~\ref{fig:velocityfield} illustrates the CO(2-1) mean velocity map and the residuals between the observed and modelled data using \textsc{3D-Barolo}. The dashed red circle (region \textit{a}), located about 2$^{\prime\prime}$ to 3$^{\prime\prime}$ northeast of the galaxy's centre, highlights an area with non-circular motion likely caused by inflows or outflows of cold molecular gas. This region, visible in the velocity map, contributes approximately 10\% of the galaxy's total flux, with about half of this flux seen in the residuals, indicating that around 5\% of the total flux is linked to non-circular motion. The estimated gas mass in this area is between \( (0.5 - 2.5) \times 10^7 \, M_{\odot} \).

We must consider that some of these residuals may be associated with extraplanar gas, often extending several kiloparsecs above the disc in spiral galaxies \citep[e.g.,][]{fraternali2002deep, li2021kinematic}. However, given the low inclination of NGC\,4593, distinguishing non-circular motions within the plane of the disc from potential outflows in the perpendicular direction is challenging. The residuals observed in our analysis are relatively small and appear consistent with non-circular motions in the plane, rather than significant extraplanar outflows. If these were true outflows, we would expect velocities much larger than the observed rotational velocity \citep{marasco2019halogas}. Additionally, extraplanar gas in other galaxies, such as the Milky Way, tends to be mostly atomic, have low densities, and constitute a small fraction of the total mass (typically less than 15\%) \citep{soding2024spatially}. Furthermore, the global SFR in NGC\,4593 does not suggest a strong star-formation-driven outflow, and the galaxy does not reside in a particularly crowded environment that would indicate gas stripping or strong interactions \citep{kormendy2006pseudobulges}.

Therefore, while we cannot completely rule out the presence of extraplanar gas, it is unlikely to be a significant factor in the observed kinematics of NGC\,4593. The presence of this excess gas flux in the residuals suggests an inflow or outflow of molecular gas. Moreover, the bump in rotation velocity, radial velocity, surface density, and velocity dispersion for the same radius away from the centre, observed in the \textsc{3D-Barolo} results (see Fig.~\ref{fig:parameters}), provides further evidence of this non-circular motion. The rotation velocity shows a noticeable rise and fall in the region, indicative of local perturbations. The radial velocity profile demonstrates deviations from the symmetric field expected for purely circular motions, suggesting inflow or outflow dynamics. An excess in surface density corresponds with the region, consistent with molecular gas accumulation. Finally, an increase in velocity dispersion in this region highlights the disturbed nature of the molecular gas. These characteristics, combined with the residual analysis, confirm the presence of non-circular motion and molecular inflows or outflows. The observed inflows or outflows and perturbations challenge the axisymmetric assumptions in the \textsc{3D-Barolo} models and underline the importance of non-axisymmetric bar potentials in shaping gas dynamics.

\subsection{Central molecular zone (CMZ) ring}

The CMZ is a dense, gas-rich region typically found in the innermost few hundred parsecs of galaxies, characterised by intense star formation and dynamic activity influenced by the gravitational effects of the galactic bar and potential AGN activity \citep{morris1983temperature, bally1987galactic}. Understanding the CMZ is crucial for studying galactic evolution due to its high concentration of molecular gas and its role as a central engine driving various dynamic processes.

Inflows of molecular gas, driven by bar-induced gravitational torques, funnel material into the CMZ, creating high-density molecular rings. These inflows are significant as they can enhance star formation rates (SFRs), fuel AGN activity, and drive turbulence within the CMZ. The interaction between inflowing gas and the central black hole (BH) can lead to episodic accretion events, contributing to AGN feedback mechanisms that regulate the growth and activity of the BH \citep{sormani2019mass, hatchfield2021dynamically, tress2024magnetic}.

Our analysis suggests the presence of a CMZ-like ring in the innermost region of NGC\,4593 (see Fig.~\ref{fig:velocityfield}). The bar of the galaxy, with a length of approximately 35$^{\prime\prime}$ \citep{Mulchaey1997}, aligns in such a way that it channels molecular gas into the CMZ-like ring through shocks and streaming motions. These patterns are consistent with bar-driven inflow mechanisms observed in other barred galaxies \citep{Athanassoula1992}. The CMZ-like ring, with its high concentration of molecular gas, could play a crucial role in the galaxy's evolution, potentially influencing SFRs, AGN activity, and the overall dynamical state of the galaxy \citep{armillotta2020life, torii2013detailed}.

We estimate that the CMZ-like ring, located in the central region with a radius of about $2^{\prime \prime}$, accounts for approximately $20\%$ of the total molecular gas mass in NGC\,4593, translating to a mass range of $\sim 0.2 - 1.0 \times 10^8 \, M_{\odot}$. The surface density of molecular gas in this CMZ-like ring is estimated to be in the range of $\Sigma \sim 0.26 - 0.34 \, \text{Jy}\,\text{km/s}$, which is significantly higher than the average surface density of molecular gas across the entire galaxy, estimated at $\Sigma \sim 0.10 - 0.14 \, \text{Jy}\,\text{km/s}$. This suggests that the surface density in the CMZ-like ring is about 2.5 times higher than the galaxy-wide average.

The velocity structure of the ring reveals strong rotational motion with signs of non-circular streaming inflow, potentially driven by bar-induced mechanisms. By examining the residual velocity field from our model, we observe significant residuals in the central region of the galaxy. The residual velocity profile indicates an inflow or outflow pattern that is perpendicular to the overall velocity map of the galaxy. Furthermore, most of these residuals at the centre are positive, suggesting an inflow of gas towards the nucleus.

\subsection{The spectral energy distribution}

To characterise galaxies and gain insights into their basic constituents, decomposing their spectral energy distributions (SEDs) into various components is a standard approach. We employed the Code Investigating GALaxy Emission (CIGALE; \citealt{boquien2019cigale, yang2020x, yang2022fitting}) to plot and analyse the SED of NGC\,4593. The photometry data, including observations and corresponding error bars used as inputs for CIGALE, are reported in Table~\ref{table:3}. For modelling the star formation history, we used a "delayed" model, characterised by a gradual rise of the star formation rate (SFR) to a maximum, followed by an exponential decline. This model can also accommodate an exponential burst to simulate a recent episode of star formation.

For the Stellar Population Synthesis, we utilised a Single Stellar Population (SSP) as defined in \citet{bruzual2003stellar}, incorporating a Chabrier Initial Mass Function, a metallicity of 0.02, and a separation age of 10 Myr. Dust attenuation was modelled using the attenuation modules from \citet{calzetti2000dust} and \citet{leitherer2002ultraviolet}, while the dust emission parameters were based on the models by \citet{dale2014two}. In addition, CIGALE allows for the inclusion of AGN emission in the SED. We employed a two-phase torus model for the AGN component, as described by \citet{stalevski2016dust}, and linked the X-ray and UV/optical emissions of the AGN using the $\alpha_{\rm ox}$–$L_{\nu,2500\mathrm{\AA}}$ relation \citep{just2007x}, with a maximum deviation of $|\Delta \alpha_{\rm ox}|_{\rm max} = 0.2$ to account for intrinsic scatter.

Initial conditions for our object were set as a face-on type-1 AGN, following the methodology outlined by \citet{mountrichas2021galaxy} and \citet{ciesla2015constraining}, who investigated type-1 and type-2 AGNs using CIGALE. Parameters such as inclination and other model settings were primarily derived from the \textsc{3D-Barolo} results presented in Section \ref{res}, while other initial conditions were allowed greater flexibility to improve the fit. This approach, while computationally demanding, yielded more precise fitting results.

\begin{figure}[!htbp]
\centering
\resizebox{\hsize}{!}{\includegraphics{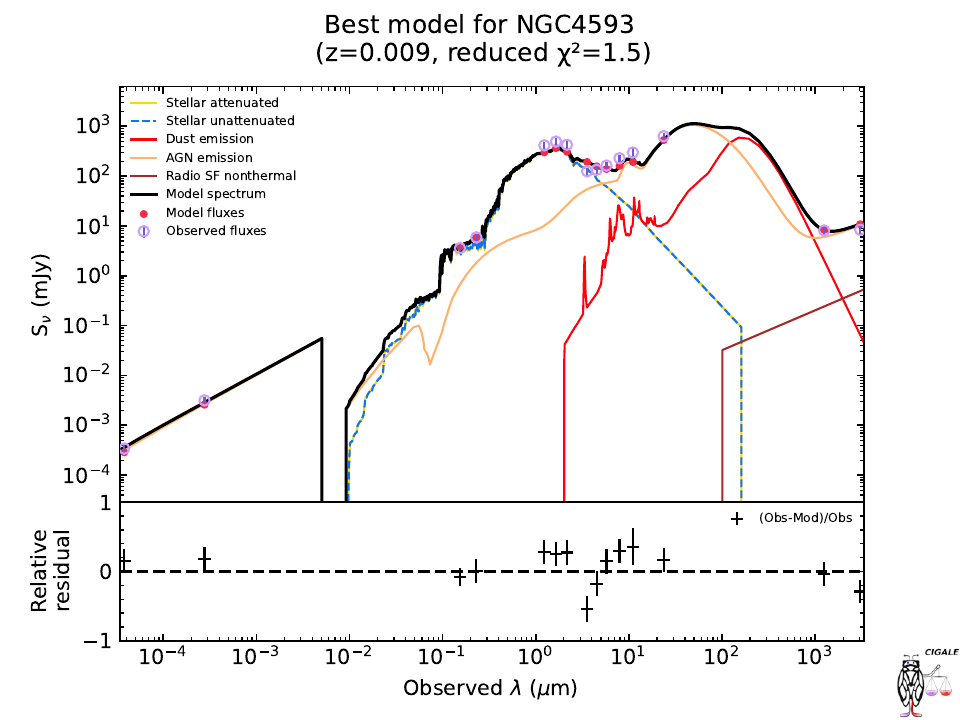}}
    \caption{SED of NGC\,4593 derived using the CIGALE software. The plot shows the observed SED, indicated by red points, compared to the best-fit model represented by the solid black line. The constituent components of the model are illustrated as follows: dust emission (red line), AGN emission (orange line), and stellar emission, both attenuated (yellow dashed line) and unattenuated (blue dashed line). The vertical axis denotes the flux density $S_{\nu}$ in mJy, while the horizontal axis represents the wavelength $\lambda$ in $\mu m$. This SED plot provides a comprehensive view of the various emission components contributing to the observed radiation from NGC\,4593.}
    \label{fig:sed}
\end{figure}

The resulting SED fit, shown in Fig.~\ref{fig:sed}, represents the best-fitting model obtained with CIGALE based on our initial conditions and photometric data. For the AGN component, specific parameters in the torus model were chosen, including an outer-to-inner radius ratio ($R_{\rm out}/R_{\rm in}$) of 20, with possible values ranging from 10 to 30. The half-opening angle was selected based on the type-1 AGN perspective, and the inclination angle was derived from the position angle (PA) of the source (see Table~\ref{tab:table1} for detailed parameter values). Although the alignment between the galaxy disc and torus may not be perfect, we assumed a standard inclination angle for consistency. Dust density gradients were set to $1.0$ for both polar angle and radial direction, ensuring a comprehensive representation of the dust distribution.

The power-law index $\delta$, modifying the optical slope of the disc, was set to $-0.36$, with an average edge-on optical depth at $9.7 \, \mu$m of $4.37 \pm 0.93$. These parameters indicate a prominent dusty torus structure around the AGN, aligned with the direct line of sight to the bright accretion disc and central engine, thus confirming the type-1 classification of NGC\,4593. Based on this model, the AGN fraction (the ratio of AGN luminosity to total FIR luminosity) was determined to be $0.876 \pm 0.052$, underscoring the significant impact of the AGN on the SED and its influence on the galaxy's overall spectrum. Detailed results derived by CIGALE are provided in Table~\ref{table:4}.

NGC\,4593 is classified as a late-type barred spiral (SB) galaxy hosting a type-1 AGN. Its bolometric luminosity ($\log(L_{\rm bol}) = 44.163$) and the substantial AGN contribution to the overall spectrum indicate a dominant AGN influence. Bright AGNs often exhibit complex interactions between SFRs and AGN luminosity, with various studies suggesting diverse relationships \citep{fanidakis2012evolution, gutcke2015sfr}. The AGN disc luminosity ($1.26 \times 10^{36} \, \text{erg/s}$) and intrinsic 2-10 keV luminosity ($L_{\nu, 2\,\text{keV}} = 1.11 \times 10^{18} \, \text{erg/s}$) support the presence of a powerful central black hole ($\log\left( \frac{M_{\rm BH}}{M_{\odot}} \right) = 6.86$).

Despite the strong AGN activity, the galaxy's SFR remains moderate. Using CIGALE, we derived an SFR of 0.43 $M_{\odot}$ yr$^{-1}$ and a stellar mass of $\log({\rm M}_*/{\rm M}_\odot) = 10.646$, consistent with previous studies. The SFR calculated from the PAH11.25$\mu$m emission \citep{mordini2021calibration} is 0.36 $M_{\odot}$/year, while that from the [CII]158$\mu$m line \citep{fernandez2020co} is 0.19 $M_{\odot}$/yr. In contrast, the total FIR luminosity estimate \citep{kennicutt1998global} provides a higher SFR of 1.2 $M_{\odot}$/yr, using the log(FIR) luminosity from \citet{spinoglio1995multiwavelength} of 43.43. The specific star formation rate (sSFR) of $4.4 \times 10^{-11} \, \text{yr}^{-1}$ indicates significant star-forming activity relative to its stellar mass, possibly influenced by AGN feedback mechanisms.

NGC\,4593 is positioned near the threshold between high-luminosity AGNs (with $\log_{10}(L_{\rm AGN}) \gtrsim 44$) and lower-luminosity AGNs. High-luminosity AGNs tend to show a strong positive correlation between SFR and $L_{\rm AGN}$, whereas lower-luminosity AGNs may exhibit a mildly negative correlation \citep{rosario2013mean}. However, studies such as those by \citet{pitchford2016extreme} and \citet{hatziminaoglou2018multiplicity} indicate that AGN luminosity and SFR are not always strongly correlated, with some FIR-bright AGNs displaying signs of companions or mergers.

Our findings place NGC\,4593 within the high-luminosity AGN regime. The AGN fraction relative to the IR luminosity is $0.87$, and its bolometric luminosity exceeds $10^{44}$ erg/s, highlighting the dominance of AGN activity in its SED. While this dominance indicates a significant AGN contribution to the galaxy's IR emission, it does not imply a straightforward correlation between AGN luminosity and star formation processes. This result suggests that while AGN activity significantly influences the galaxy's infrared emission, it may not directly drive star formation.

According to \citet{zhuang2023evolutionary}, AGNs and their host galaxies evolve along specific trajectories on the black hole mass (M$_{\rm BH}$) versus stellar mass (M$_*$) plane. Galaxies above the M$_{\rm BH}$-M$_*$ relation typically grow horizontally with substantial stellar mass increases, while those on the relation evolve proportionally in both M$_{\rm BH}$ and M$_*$. Our analysis reveals that NGC\,4593 aligns with this "horizontal path," suggesting that radiative-mode feedback cannot fully suppress star formation in such AGNs, and kinetic-mode feedback is insufficient to halt long-term star formation. The high AGN fraction and substantial stellar mass growth support the conclusion that radiative-mode feedback does not effectively constrain star formation in NGC\,4593.

\subsection{Challenges in kinematic modelling}
Our analysis reveals a significant challenge in modelling the dynamics of NGC\,4593 due to the clear presence of an m=1 feature, which indicates lopsidedness that the \textsc{3D-Barolo} model, which imposes axisymmetry, struggles to account for. This discrepancy suggests potential variations in fits between the northern and southern regions of the galaxy, which the model cannot adequately address. The proximity of the bar's PA to the kinematic PA complicates the determination of true inclinations, potentially leading to systematic biases, especially in the presence of an outer ring, which is rarely circular. The assumption of circular orbits by \textsc{3D-Barolo}, when the orbits may actually be elongated, introduces further inaccuracies in the modelled inclination and fails to capture the lack of a velocity gradient in the modelled PV diagrams. Consequently, the radial velocities predicted by \textsc{3D-Barolo} cannot accurately reproduce the observed elongated orbits, as they assume isotropic motion, overlooking the complexities of elliptical orbits that do not conform to simple inflow or outflow patterns.

A critical limitation in our modelling is that the CO(2-1) line does not represent the entire molecular gas content, and relying solely on it could lead to an underestimation of the total molecular gas mass. Furthermore, the interferometric observations used in this study filter out extended emissions, resulting in missing flux, suggesting that the actual molecular gas mass could be higher than our model predicts. 

This missing flux is particularly important in regions with significant non-circular motions, such as the areas affected by the bar and spiral arms, where the model’s limitations in capturing the full extent of gas dynamics are most evident. Additionally, the \textsc{3D-Barolo} model, with its inherent assumption of axisymmetry, cannot capture the entire molecular gas distribution, leading to the omission of the single arm of the galaxy in our analysis. This limitation underscores the need for a more comprehensive modelling approach to fully understand the gas dynamics in NGC\,4593.

\begin{table}[h!]
\centering
\caption{CIGALE-derived parameters for NGC\,4593}
\label{table:4}
\begin{tabular}{l l}
\hline\hline
Parameter & Value \\
\hline
$L_{\mathrm{IR,AGN}} / L_{\mathrm{IR,total}}$ & $0.876 \pm 0.052$ \\
$\log(L_{\mathrm{AGN,torus}}\, [\mathrm{erg\,s^{-1}}])$ & $36.193 \pm 0.049$ \\
$\log(L_{\mathrm{AGN,disc}}\, [\mathrm{erg\,s^{-1}}])$ & $36.096 \pm 0.116$ \\
$\log(L_{\mathrm{AGN,total\,dust}}\, [\mathrm{erg\,s^{-1}}])$ & $36.510 \pm 0.050$ \\
$\log(L_{\mathrm{AGN,6\,\mu m}}\, [\mathrm{erg\,s^{-1}}])$ & $35.784 \pm 0.048$ \\
$\log(L_{\mathrm{AGN,accretion\,power}}\, [\mathrm{erg\,s^{-1}}])$ & $36.492 \pm 0.047$ \\
$\log(L_{\mathrm{AGN,polar\,dust}}\, [\mathrm{erg\,s^{-1}}])$ & $36.224 \pm 0.057$ \\
$\log(L_{\mathrm{AGN}}\, [\mathrm{erg\,s^{-1}}])$ & $36.651 \pm 0.048$ \\
$\log(L_{\mathrm{stellar}}\, [\mathrm{erg\,s^{-1}}])$ & $37.402 \pm 0.035$ \\
$\log(M_*/M_\odot)$ & $10.633 \pm 0.039$ \\
$\log(M_{\mathrm{stellar\,gas}}\, [M_\odot])$ & $10.535 \pm 0.041$ \\
$\log(\mathrm{SFR}_{\mathrm{integrated}}\, [M_\odot\,\mathrm{yr}^{-1}])$ & $10.888 \pm 0.039$ \\
$\log(\mathrm{SFR}_{\mathrm{current}}\, [M_\odot\,\mathrm{yr}^{-1}])$ & $-0.367 \pm 0.045$ \\
$\log(\mathrm{SFR}_{100\,\mathrm{Myr}}\, [M_\odot\,\mathrm{yr}^{-1}])$ & $-0.329 \pm 0.042$ \\
$\log(\mathrm{SFR}_{10\,\mathrm{Myr}}\, [M_\odot\,\mathrm{yr}^{-1}])$ & $-0.364 \pm 0.045$ \\
\hline
\end{tabular}
\tablefoot{
CIGALE-derived parameters for NGC\,4593, including luminosity components of the stellar population, active galactic nucleus (AGN), accretion disc, and dust structures (polar, torus, and total). SFRs reflect the current and past 10 and 100 million-year averages. These values are estimated based on CIGALE fitting results.
}
\end{table}

\section{Summary and conclusions}

We used CO(2-1) observations from ALMA to study the kinematics and morphology of molecular gas in the Seyfert 1 AGN galaxy NGC\,4593, which is a barred spiral galaxy. For this work, we used \textsc{3D-Barolo} and \textsc{discFit} to analyse the gas kinematics and morphology. Additionally, we utilised multi-wavelength data to construct the SED of the galaxy using CIGALE. Our analysis revealed that the molecular gas mass in NGC\,4593 is in the range \( 1.0 \sim 5.0 \times 10^8 \, M_{\odot} \), derived from the CO(1-0) line luminosity from the CO(2-1) using an excitation factor typical of active galaxies \citep{kamenetzky2016relations}.

The gas distribution exhibits a one-arm structure, indicative of an \( m = 1 \) density wave or perturbation. Such a feature, rare in the nuclear regions of barred galaxies, suggests a dominant regular velocity field, implying limited non-circular motion. Yet, the evident \( m = 1 \) feature challenges the axisymmetry assumptions inherent in the \textsc{3D-Barolo} models.

The mass of the central SMBH was estimated based on the relationship between the CO line width (\(W_{50}\)) and SMBH mass. Our analysis yielded a BH mass of \(\log \left( \frac{M_{\text{BH}}}{M_{\odot}} \right) = 6.89 \pm 0.04\), which aligns with previous estimates from reverberation mapping.

In the region about 2$^{\prime\prime}$ to 3$^{\prime\prime}$ ($\sim$ 220 pc) northeast of the galaxy centre (see Fig.~\ref{fig:velocityfield}), we identified significant non-circular motion likely due to molecular gas outflow. This region contains between \( (0.5 - 2.5) \times 10^7 \, M_{\odot} \) of gas, representing roughly 10\% of the total galaxy flux. Perturbations in rotation velocity, radial velocity, surface density, and velocity dispersion confirm the presence of significant non-circular motion and molecular outflows. These findings align with observations by \citet{barbosa2006gemini}, who noted similar perturbations in ionised gas kinematics.

The inner structure of NGC\,4593 appears to be shaped by a CMZ-like ring in the innermost region of the galaxy. This ring, which could account for up to 20\% of the total molecular gas mass, exhibits strong rotational motion with signs of non-circular inflow patterns. The bar-driven inflow mechanisms channel molecular gas into this CMZ-like ring, thereby affecting star formation and AGN activity in the central region. This prominent bar structure significantly influences the gas dynamics.

The SED fitting using CIGALE confirmed the presence of a strong AGN component in NGC\,4593. The AGN fraction to IR luminosity is $0.87$, underscoring the significant impact of the AGN on the galaxy's overall spectrum. The inferred star formation rate is not large, which might suggest the suppression of star formation processes because of AGN quenching. NGC\,4593 fits well within the high-luminosity AGN regime, with a bolometric luminosity above \(10^{44}\) erg/s. The galaxy's stellar mass growth suggests that radiative-mode feedback is ineffective in fully suppressing star formation. Instead, the AGN activity remains tightly linked with star formation, consistent with the coevolution trajectory outlined by \citet{zhuang2023evolutionary}. 

NGC\,4593 exhibits complex gas dynamics due to the interplay between bar dynamics and AGN activity. The presence of a CMZ-like ring and significant molecular outflows highlight the impact of non-axisymmetric bar potentials on shaping gas dynamics.


\begin{acknowledgements}
This paper makes use of data from ALMA program ADS/JAO.ALMA\#2017.1.00236.S and ADS/JAO.ALMA\#2018.1.00576.S. ALMA is a partnership of ESO (representing its member states), NSF (USA) and NINS (Japan), together with NRC (Canada) and NSTC and ASIAA (Taiwan) and KASI (Republic of Korea), in cooperation with the Republic of Chile. The Joint ALMA Observatory is operated by ESO, AUI/NRAO and NAOJ. KK acknowledges the support of the European Southern Observatory (ESO) through the SSDF grant, which facilitated a visit to ESO contributing to this work.
 CR acknowledges support from Fondecyt Regular grant 1230345 and ANID BASAL project FB210003.
 JAFO acknowledges financial support by the Spanish Ministry of Science and Innovation (MCIN/AEI/10.13039/501100011033), by ``ERDF A way of making Europe'' and by ``European Union NextGenerationEU/PRTR'' through the grants PID2021-124918NB-C44 and CNS2023-145339; MCIN and the European Union -- NextGenerationEU through the Recovery and Resilience Facility project ICTS-MRR-2021-03-CEFCA. MPS acknowledges support from grants RYC2021-033094-I and CNS2023-145506 funded by MCIN/AEI/10.13039/501100011033 and the European Union NextGenerationEU/PRTR.
PA warmly thanks the Department of Physics, Section of Astrophysics, Astronomy and Mechanics of the Aristotle University of Thessaloniki (Greece),  the Institute of Theoretical Astrophysics, University of Oslo (Norway), and the Fukui University of technology, Fukui (Japan) for their hospitality when part of this work was written. 
\end{acknowledgements}

%
%

\bibliographystyle{aa} 
\bibliography{final}
\appendix
\end{document}